%% file: main.tex
\documentclass{aa}
\usepackage{amssymb,amsfonts,amsmath,times,pict2e}
\usepackage{natbib}
\usepackage{graphicx}
\usepackage{enumerate}
\usepackage{subfigure}
\usepackage{bm}
\usepackage{amsmath,amsbsy}
\usepackage{underscore}
\usepackage{float}

\usepackage{color}
\usepackage{ulem}
\definecolor{mygray}{gray}{0.6}
\definecolor{magenta}{rgb}{0.858, 0.188, 0.478}

\newcommand{\Rmnum}[1]{\expandafter\@slowromancap\romannumeral #1}

\newcommand{\ccc}[1]{\textcolor{red}{[{\small #1}]}}

\usepackage{xspace}
\newcommand{\fg}[1]{Fig.~\ref{fig:#1}}
\newcommand{\Fg}[1]{Figure~\ref{fig:#1}}

\newcommand{\eq}[1]{Eq.~(\ref{eq:#1})\xspace}
\newcommand{\Eq}[1]{Equation~(\ref{eq:#1})\xspace}
\newcommand{\eqs}[2]{Eqs.\ (\ref{eq:#1}) and (\ref{eq:#2})}

\newcommand{\tb}[1]{Table~\ref{tab:#1}\xspace}
\newcommand{\Tb}[1]{Table~\ref{tab:#1}\xspace}
\newcommand{\se}[1]{Sect.~\ref{sec:#1}\xspace}
\newcommand{\ses}[2]{Sects.\ \ref{sec:#1} and \ref{sec:#2}}

\newcommand{\ie}{i.e.}

\newcommand{\eg}{e.g.}

\newcommand{\AU}{ \  \rm AU}

\newcommand{\Ms}{ \  \rm M_\odot }

\newcommand{\yr}{ \   \rm yr}

\newcommand{\Me}{ \ \rm M_\oplus}

\newcommand{\taus}{ \tau_{\rm s}}

\newcommand{\vk}{v_{\rm K}}
\newcommand{\Omegak}{\Omega_{\rm K}}

\newcommand{\Sigmap}{\Sigma_{\rm P}}
\newcommand{\Sigmag}{\Sigma_{\rm gas}}
\newcommand{\hg}{h_{\rm gas}}
\newcommand{\Mp}{M_{\rm p}}
\newcommand{\vhw}{v_{\rm hw}}
\newcommand{\rp}{r_{\rm p}}
\newcommand{\taue}{\tau_{\rm e}}
\newcommand{\taum}{\tau_{\rm m}}

\newcommand{\qp}{q_{\rm p}}
\newcommand{\qc}{q_{\rm hw/sh}}
\newcommand{\ep}{e_{\rm p}}

\newcommand{\changed}[1]{#1}

\usepackage{CJKutf8}

\begin{document}

\title{Catching drifting pebbles }
\subtitle{I. Enhanced pebble accretion efficiencies for eccentric planets }

\author{ Beibei Liu  \inst{1}, Chris W. Ormel\inst{1} }
\institute{Anton Pannekoek Institute (API), University of Amsterdam, Science Park 904,1090GE Amsterdam, The Netherlands\label{inst1}\\
\email{b.liu@uva.nl,c.w.ormel@uva.nl}
 }

\date{\today}


\input{abstract.tex}

{}
{}
{}
{}
{}


\maketitle

\input{introduction.tex}

\input{method.tex}

\input{results.tex}

\input{application.tex}

\input{conclusion.tex}

\input{appendix.tex}

\begin{acknowledgements}
We thank Anders Johansen, Carsten Dominik for useful discussions, and Ramon Brasser for proofreading the manuscript. 
We also thank the anonymous referee for their insightful suggestions and comments. 
 B.L.\ and C.W.O\ are supported by the Netherlands Organization for Scientific Research (NWO; VIDI project 639.042.422).  
\end{acknowledgements}

\bibliographystyle{aa}
\bibliography{reference}

\end{document}

%% file: abstract.tex
\abstract{
    Coagulation theory predicts that micron-sized dust grains grow into pebbles, which drift inward towards the star when they reach sizes of mm$-$cm. When they cross the orbit of a planet, a fraction of these drifting pebbles will be accreted. In the pebble accretion mechanism, the combined effects of the planet's gravitational attraction and gas drag greatly increase the accretion rate.} 
    {We calculate the pebble accretion efficiency $\varepsilon_\mathrm{\rm 2D}$ --  the probability that a pebble is accreted by the planet -- in the 2D limit (pebbles reside in the midplane).  In particular, we investigate the dependence of $\varepsilon_{\rm 2D}$ on the planet eccentricity and its implications for planet formation models.}
{We conduct N-body simulations to calculate the pebble accretion efficiency in both the local frame and the global frame. With the global method we investigate the pebble accretion efficiency when the planet is on an eccentric orbit.  
}
{We find that the local and the global methods generally give consistent results. However, the global method becomes more accurate when the planet is more massive than a few Earth masses or when the aerodynamic size (Stokes number) of the pebble is larger than $1$.
  The efficiency increases with the planet's eccentricity once the relative velocity between the pebble and the planet is determined by the planet's eccentric velocity. At high eccentricities, however, the relative velocity becomes too high for pebble accretion. The efficiency then drops significantly and the accretion enters the ballistic regime. We present general expressions for $\varepsilon_\mathrm{2D}$. Applying the obtained formula to the formation of a secondary planet, in resonance with an already-formed giant planet, we find that the embryo grows quickly  due to its higher eccentricity.  
}
{ The maximum $\varepsilon_\mathrm{2D}$ for a  planet on an eccentric orbit is several times higher than for a planet on a circular orbit, but this  increase gives  the planet an important headstart and boosts its following mass growth.   The recipe for $\varepsilon_\mathrm{2D}$ that we have obtained  is designed to be implemented into N-body codes to  simulate the growth and evolution of planetary systems.   
 }

%% file: introduction.tex
\section{Introduction}
\label{introduction}

In protoplanetary disks, micron-sized dust grains grow by coagulation \citep{Weidenschilling1993,Dominik2007,Johansen2014}. Typically, dust coagulation can be divided into two phases: an initial growth phase, where particles grow from micron-sized dust grains into much larger pebbles; and a drift phase, where particles are transported to the inner disk regions. Aerodynamically, a particle is defined in terms of a dimensionless quantity $\taus$ (often referred to as the Stokes number). Dust grains have $\tau_s$$\ll$$10^{-3}$ and are strongly coupled to the gas. Particles that have become aerodynamically larger ($\taus\sim10^{-2}-1$), on the other hand, decouple from the gas and  drift inward \citep{Weidenschilling1977a}. The radial drift effectively limits the size of the particles. Other processes that limit the size occur once the particles' relative velocities reach the bouncing or fragmentation thresholds \citep{Guttler2010}.  The drifting particles are defined as  pebbles, whose size therefore depends on the material strength and the disk properties  but typically lies in the mm$-$cm range \citep{Brauer2008,Birnstiel2010}.  Because  the growth timescale of pebbles  increases  with the disk radius, the dust evolution proceeds  in an inside-out manner. The interior, planet-forming regions are therefore constantly supplied by pebbles that drift from the outer regions of disks \citep{Birnstiel2012,Lambrechts2014,Sato2016,Krijt2016}.

Observationally, the existence of pebble-sized particles in disks is inferred from optically thin emission at sub-mm and radio wavelengths. Particles that have grown to sizes larger than the observing wavelength tend to emit as a gray body. Therefore,  a millimeter spectral index that is reduced compared to the interstellar medium (ISM) value is a signature of the existence of pebble-sized particles \citep{Natta2004,Draine2006,Perez2015,Tazzari2016}.
Assuming a temperature (profile) and a millimeter opacity (quantities that are rather uncertain), the total mass of the pebble reservoir can be calculated \citep{Williams2011}. Typical values for the pebble disk mass lie from a few Earth masses to hundreds of Earth masses \citep{Ricci2010,Andrews2013,Ansdell2016,Barenfeld2016,Pascucci2016}. 
\changed{However,  the observed pebble mass may not  represent the full reservoir of solid material in the disk. For instance,
depending on the rapidity of the planet formation process, some of the erstwhile pebbles may already have been locked up in larger bodies, invisible to millimeter disk observations.} Indeed, \cite{Ansdell2017} find lower average dust masses for the older $\sigma$ Orionis cluster.

When the drifting pebbles cross the orbit of  big bodies (planetesimals, planetary embryos, or planets), a fraction of them can be accreted by the combined effects of the planet's gravitational attraction and gas drag \citep{Ormel2010,Lambrechts2012}. This process is known as pebble accretion (see \cite{Ormel2017,Johansen2017} for recent reviews). Depending on the importance of gas drag during the pebble-planet encounter, the accretion can be classified into two regimes: settling and ballistic. In the classical, planetesimal-driven planet formation paradigm \citep{Safronov1972,Lissauer1987,Ida2004a}, accretion  takes place entirely through ballistic interactions. In the ballistic regime, gas drag is unimportant during the encounters and accretion relies on hitting the surface. On the other hand, in the settling regime, pebbles settle onto the planet at their terminal velocity (the velocity where the planet's gravity balances gas drag). The accretion rate does not depend on the planet's physical radius; only the mass of the planet matters. For large planets accreting $\taus\sim1$ particles, the accretion cross section can be as large as the planet's Hill sphere.

The large accretion cross sections and continuous supply of pebbles from the outer disk may, at first sight, offer ideal conditions to produce planets. However, there is one catch: the planet may not accrete all pebbles, simply because  they drift too fast or they are stirred to a large height. The viability of pebble accretion as a planet formation process therefore depends not only  on the (large) accretion cross sections, but also on the disk conditions, i.e., whether pebbles drift fast and how settled they are. Specifically, we define the pebble accretion efficiency ($\varepsilon$) as the number of pebbles that are accreted over the total number of pebbles that the disk supplies. In terms of mass fluxes, the definition reads \footnote{This definition is the same as the filtering factor/efficiency of \changed{\cite{Guillot2014} and \cite{Lambrechts2014}}.}
\begin{equation}
    \varepsilon \equiv \frac{\dot{M}_\mathrm{PA}}{\dot{M}_\mathrm{disk}},
\end{equation}
where $\dot{M}_\mathrm{PA}$ is the pebble accretion rate on the planet and $\dot{M}_\mathrm{disk}$ is the mass flux of pebbles through the disk.  Put simply, $\varepsilon$ is the probability that a pebble is accreted by the planet. A high value of $\varepsilon$ indicates that pebble accretion is efficient. For example, if $\varepsilon=1$ (the highest value possible) and the disk contains a total $10\Me$ in pebbles, an initial $1\Me$ planet will grow to $11\Me$, which is large enough to trigger giant planet formation \citep{Pollack1996}. On the other hand, when only one in a  thousand pebbles is accreted ($\varepsilon=10^{-3}$), the gained planetary mass will be $10^{-2}\Me$, meaning that the planet growth has significantly stalled. The efficiency $\varepsilon$ is a \textit{local} quantity: it can be computed entirely from the conditions at the location of the planet. Nevertheless, as illustrated by the example, $\varepsilon$ carries global significance since it directly states the total amount of pebbles a disk needs to contain in order to grow planets, \changed{i.e., how efficiently the pebble mass can be converted into planet mass}.

The goal of our  paper series is to obtain a general recipe for pebble accretion efficiency $\varepsilon$, \ie, to quantify how $\varepsilon$ depends on planet properties (e.g., the planet mass, the eccentricity and the inclination), pebble properties (the pebble's aerodynamical size $\taus$), and disk properties (temperature and turbulence strength).  In this work (hereafter Paper I) only the planar limit is considered, where we assume that the pebbles have settled to the disk midplane, or more generally, that the pebble accretion radius exceeds the  scaleheight of the pebble layer. We elucidate the role of the planet eccentricity on the pebble accretion efficiency in this 2D limit ($\varepsilon_{\rm 2D}$).  In the subsequent paper (Paper II, \cite{Ormel2018}), we calculate the 3D pebble accretion efficiency ($\varepsilon_\mathrm{3D}$) by investigating the roles of the planet inclination and the disk turbulence.  The prescriptions that we obtain in these studies can be  implemented in N-body codes, in order to study the formation and long-term dynamical evolution of planetary systems.

In the literature, most theoretical and numerical work consider pebble accretion for planets on circular orbits \citep{Ormel2010,Lambrechts2012,Lambrechts2014,Guillot2014,Morbidelli2015,Bitsch2015b,Levison2015a,Ida2016,Matsumura2017,Picogna2018}. 
The relatively velocity between the pebble and the planet changes when the planet is on an eccentric orbit. Therefore the accretion efficiency can change as well. During planet formation, planets can acquire  moderate eccentricities through mechanisms as planet-planet scattering, secular and mean motion resonances \citep{Lee2002,Raymond2006,Zhou2007b,Zheng2017}. To investigate how the accretion efficiency depends on the planet eccentricity is therefore relevant.  \cite{Johansen2015} already considered the eccentric situation in which they focus on planetesimals with relatively low eccentricities in the local (co-moving) frame. In this paper, we will also conduct numerical orbital integrations in the global (heliocentric) frame, which is especially appropriate for planets with high eccentricities.
        


The paper is structured as follows. In \se{method}, we outline two approaches to calculate $\varepsilon_{\rm 2D}$ by considering the equation of motion in the local frame and the global frame, respectively. Results are presented in \se{results}. We compare the results from the above two approaches (\se{compare}), and investigate how $\varepsilon_{\rm 2D}$ depends on properties of the planet, the pebble and the disk (\se{ecc}).
We also provide analytical fit expressions for $\varepsilon_{\rm 2D}$ (\se{expression}).  
In \se{application}, we apply our results by assessing how fast a secondary planet can grow from a planetary embryo, in the presence of an already-formed giant planet. We summarize our key findings in \se{conclusion}. The derivation of the pebble accretion efficiency expressions and list of notations are given in the Appendix A.

%% file: method.tex
\section{Method}
\label{sec:method}

In this section, we present two ways to calculate the pebble accretion efficiency. One approach is to treat the pebble's motion with respect to the planet in a non-inertial, local frame (\se{local}).  The alternatively approach is to consider it in a global frame centred on the star (\se{global}). Three numerical methods based on the above two approaches are described in \se{integration}.
\subsection{Local frame}
\label{sec:local}
\begin{figure*}[tbh!]
      \includegraphics[scale=0.65, angle=0]{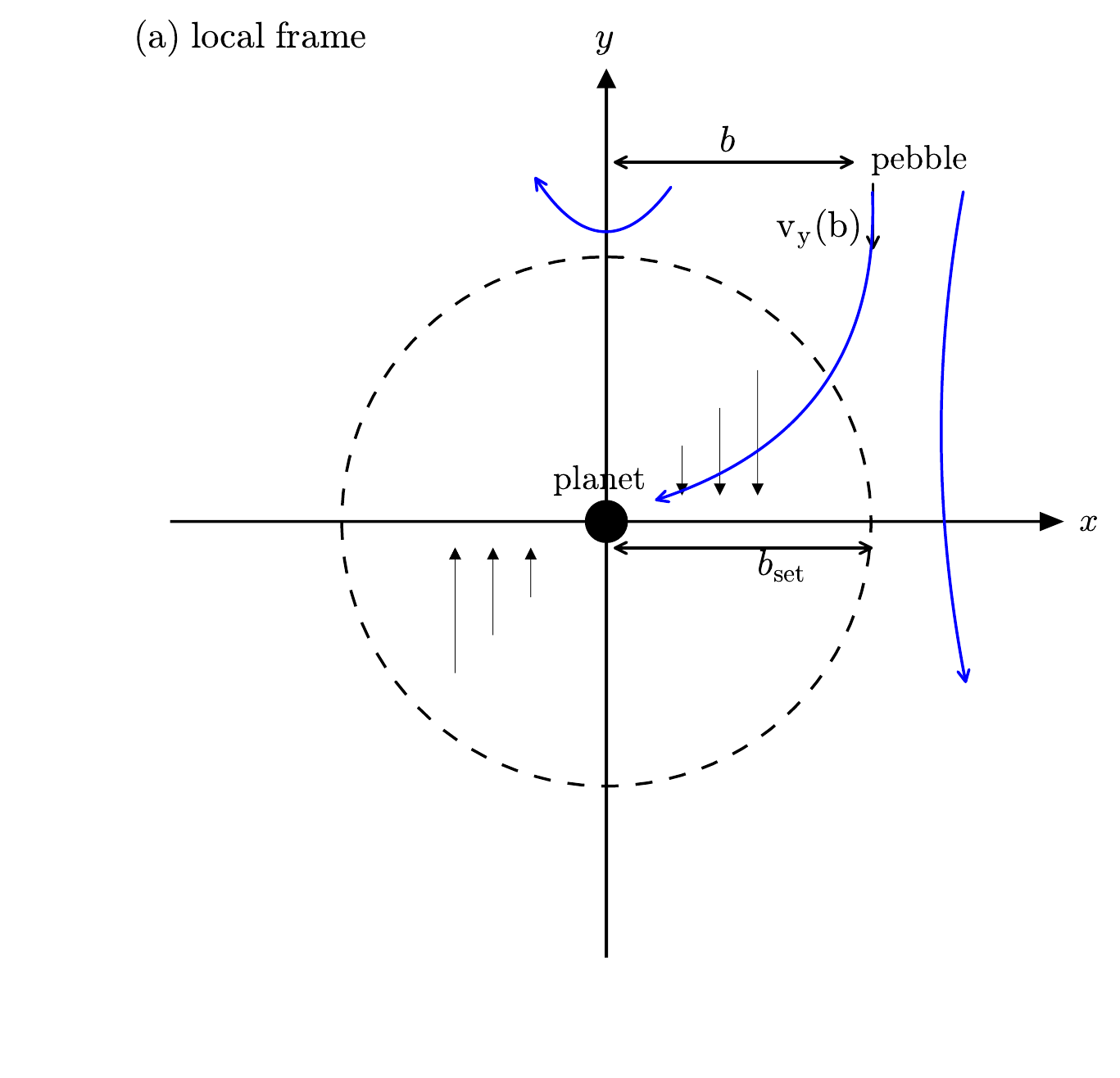}
       \includegraphics[scale=0.65, angle=0]{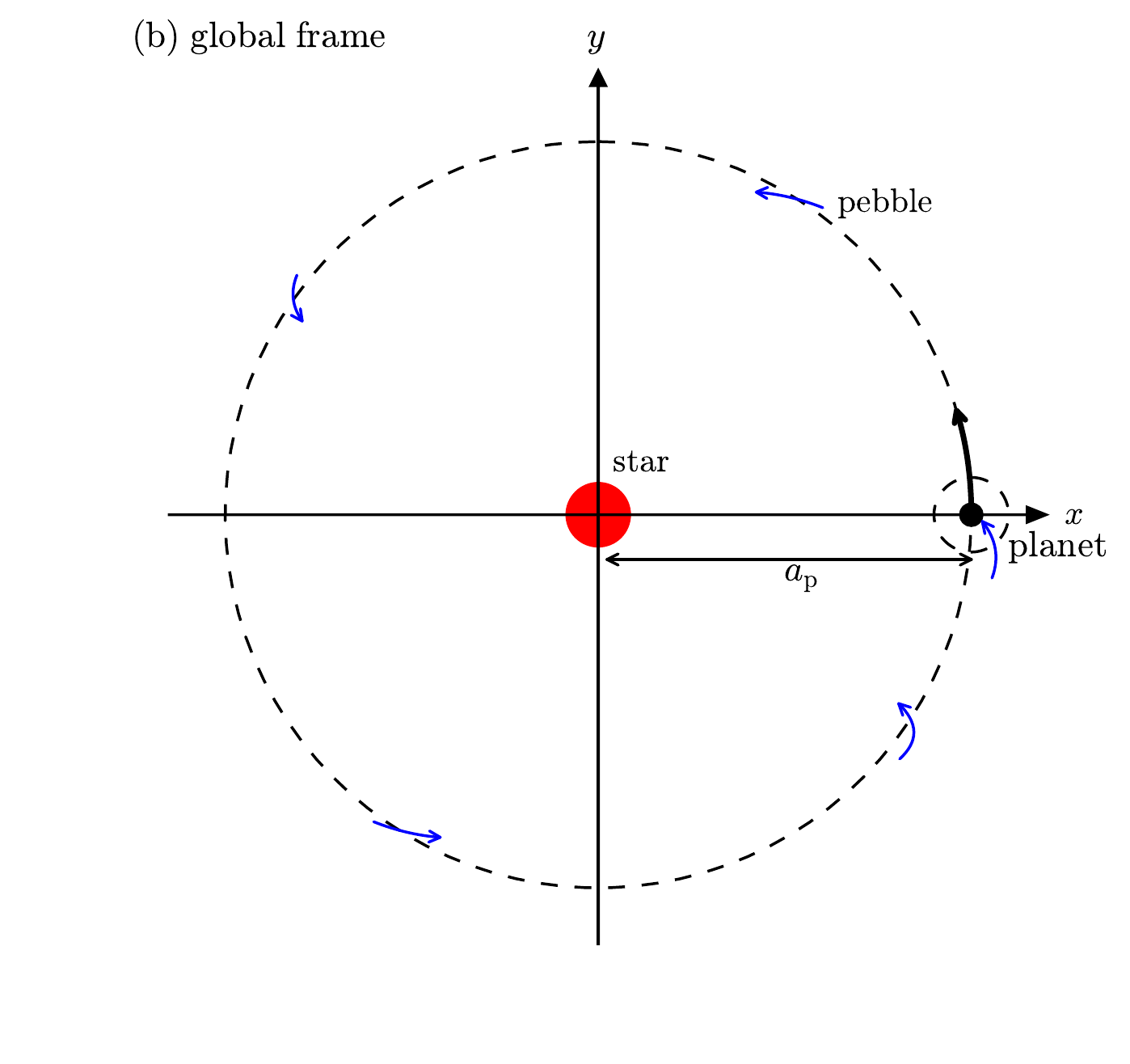}
       \caption{
      The sketch illustrating pebble accretion in two different frames. 
      a) Local frame: the co-moving frame is centred on the planet where the x-axis is pointing radially away from the star and the y-axis follows the orbital direction of the planet. Grey arrows purely indicate the gas velocity of the shearing flow, and the blue arrow shows the trajectory of the pebble, with the x-axis impact distance $b$ and the y-axis velocity $v_{\rm y}(b)$.   
      b) Global frame: the planet orbits around the central star with a semi-major axis $a_{\rm p}$.  The black arrow depicts the planet motion and  the blue arrows illustrate the trajectories of pebbles that cross the orbit of the planet. } 
\label{fig:frames}
\end{figure*} 

In literature, \cite{Ormel2010} and \cite{Lambrechts2012} adopt a local frame approach to investigate the pebble-planet interaction. As shown in \fg{frames}a, the coordinate is in a comoving frame centred on and rotating with the planet. In the shearing box approximation, the pebble's motion is linearized with respect to the planet (Eq. (7) in \cite{Ormel2010}).

In a $2$D limit, the pebble accretion efficiency in the local frame is given by \citep{Guillot2014,Ormel2017}
\begin{equation}
    \varepsilon
    = \frac{\dot{M}_\mathrm{PA}}{\dot{M}_\mathrm{disk}}
    = \frac{\int  |v_{\rm y} (b)| \bar{H}(b)  \Sigmap  {\rm d} b  }{2\pi r_{\rm p} v_{\rm r} \Sigmap },
    \label{eq:eta-local}
\end{equation}
where  $b$ is the pebble's impact distance measured from the $x$-axis and $v_{\rm y} (b)$ is the pebble's velocity perpendicular to $b$  (\fg{frames}a), $ \Omega_{\rm p} $ is the angular velocity of the planet, and $\bar{H}(b)$ is the Heaviside step function; $\bar{H}(b) = 1$ means the pebble with a impact distance $b$ hits the planet, otherwise, $\bar{H}(b)= 0$. 
The total pebble mass flux is given by $\dot{M}_\mathrm{disk} = 2\pi  \rp v_{\rm r} \Sigmap$, where $v_{\rm r}$ is the radial drift velocity of the pebble and $\Sigmap$ is the pebble surface density. All above quantities refer to  the values at the planet's location ($r_{\rm p}$).
In \eq{eta-local}, the efficiency is  obtained by the mass flux ratio of the accreted pebbles to total pebbles that cross the orbit of the planet. 

\subsection{ Global frame}
\label{sec:global}

Here we introduce a new method to calculate the pebble accretion efficiency $\varepsilon$. In the global frame centred on the central star,  we integrate the equations of motion to find the orbital evolution of the planet as well as pebbles. As pebbles drift in from regions exterior to the planet and bypass its orbit,  the accretion efficiency is simply obtained by counting the fraction of pebbles that hit the planet.

The equation of motion for the pebble in heliocentric coordinates is
\begin{multline}
    \frac{\mathrm{d}^2 \bm{ r }}{\mathrm{d} t^2}
  = \bm{F_{\star}} + \bm{F}_{\rm planet} + \bm{F}_{\rm drag}\\
  =  - GM_{\star} \frac{  \bm{r} }{r^3}   + GM_{ \mathrm{p}} \left( \frac{\bm{r}_{ \rm p} - \bm{ r}}{ | \bm{r}_{\rm p} - \bm{ r} |^3 }  -\frac{\bm{r}_{\rm p}}{r_{\mathrm p}^3}  \right)  -\frac{  \bm{v} - \bm{ v}_{\rm gas}}{t_{\rm s}}, 
  \label{eq:eq-pebble}
\end{multline}
where $G$ is the gravitational constant, $\bm{r} = (x,y,z)$, $\mathbf{r_{\rm p}} = (x_{\rm p},y_{\rm p},z_{\rm p})$  are the positions of the pebble and the planet with respect to the central star, $M_{\star}$ and $M_{\rm p }$ are the masses of the central star and the planet, $\bm{v} = (v_{\rm x},v_{\rm y},v_{\rm z})$ and $\bm{v_{\rm gas }} = (v_{\rm K}-v_{\rm hw})\mathbf{e}_{\rm \phi}$  are the velocity vector  of the pebble and the gas's azimuthal velocity at the pebble's location $r$,  and $t_{\rm s}$ is the pebble's stopping time.
It should be note that $v_{\rm hw} = \eta v_{\rm K} $ is the headwind velocity which measures the deviation between the Keplerian velocity ($v_{\rm K}$) and the gas azimuthal velocity, $\eta = -\frac{1}{2}  \frac{H_{\rm gas}^2 \partial \mathrm{log} P}{r^2 \partial \mathrm{log} r}$ is the headwind prefactor,  $H_{\rm gas}$ is the disk scale height and $P$ is the gas pressure at the planet's location $\rp$.

In \eq{eq-pebble}, the total (per unit mass) forces of the pebble include the gravitational forces from the central star and the planet, and the drag force from the disk gas.

The pebble is treated as a  body with zero gravitational mass. The planet only feels the gravity from the central star, and its equation of motion is expressed as   
 \begin{equation}
  \frac{\mathrm{d}^2 \bm{ r_{\rm p} }}{\mathrm{d} t^2} =   -  \frac{ GM_{\star} \bm{r_{\rm p}} }{r_{\rm p}^3}. 
  \label{eq:eq-planet}
\end{equation}
Thus, we can obtain the orbital motions of the planet and the pebble from \eqs{eq-pebble}{eq-planet}.

Integrating \eq{eq-pebble} in the global frame is computationally expensive, in particular when the pebble is strongly coupled to the gas (small $t_s$), resulting in a tiny relative velocity  ($\Delta v = v- v_{\rm gas}$). 
In such cases, the pebble tends to move at its terminal velocity, where gas drag balances the gravitational forces. Therefore, the relative acceleration $\frac{\mathrm{d}^2}{\mathrm{d}t^2}  (\bm{r}-\bm{r_{\rm gas}})$ vanishes and the pebble's velocity can be approximated as       
\begin{equation}
    \bm{v} \approx   \bm{v_{\rm gas}} + t_{\rm s} \left( \bm{F_{\star} + \bm{F_{\rm planet}} - \Omega_{\rm gas }^2 r }  \right),
      \label{eq:sc}
\end{equation}
We refer to this approximation as the strong coupling approximation (SCA) \changed{ \citep{Johansen2005}}.  This condition is satisfied when the pebble and the gas are well coupled ($\bm{v}\approx \bm{v}_{\rm gas}$). It is invalid when the perturbation from the planet is  significant or $\taus$ becomes large.
Clearly, in \eq{sc} the velocity can be explicitly calculated  whereas in \eq{eq-pebble} the velocity needs to solve from a differential equation. Therefore, the SCA  greatly simplifies the calculation of velocity compared to directly integrating the equation of motion.

The efficiency in the global frame is simply
\begin{equation}
    \varepsilon_{\rm 2D}
    = \frac{N_{\rm hit}}{N_{\rm tot}},
        \label{eq:eta-global}
\end{equation}
 where $N_{\rm tot}$ is total number of pebbles across the orbit of the planet, and  $N_{\rm hit}$ is the number of pebbles accreted by the planet.

\subsection{Three numerical methods}
\label{sec:integration}
We employ three methods to calculate the pebble accretion efficiency:
\begin{enumerate}
    \item \textit{Local} -- direct integration of the equation of motion \citep{Ormel2010} in the local frame. The pebble accretion efficiency $\varepsilon_{\rm 2D}$ is obtained from \eq{eta-local};
    \item \textit{Global direct} -- direct integration of the equation of motion in the global frame (\eq{eq-pebble}), and $\varepsilon_{\rm 2D}$ is calculated directly from the fraction of pebbles that hit the planet (\eq{eta-global});
    \item \textit{Global hybrid} -- The SCA (\eq{sc}) is applied for $|r_{p}- r|$$>$$2 R_{\rm H}$; otherwise switch to direct integration (\eq{eq-pebble}). The efficiency $\varepsilon_{\rm 2D}$ is also obtained the same as  global direct method.  The Hill radius is defined as $R_{\rm H} \equiv (M_{\rm p}/3M_{\star})^{1/3}a_{\rm p}$ and $a_{\rm p}$ is the planet's semi-major axis. 
\end{enumerate}

\changed{We assume that the inclination of the planet $i_{\rm p} $ is zero. More generally, the inclination can be neglected when it is much smaller than the the scale height of the pebble disk.  In this work we only consider the $2$D planar  limit where the planet and pebbles are all settled into the disk midplane.  3D effects (planet inclination and disk turbulence) will be modeled in detail in paper II}.  
 The local method can be used for a planet on a circular orbit ($\ep=0$). To model a planet on an eccentric orbit in the local frame, the elliptic motion of the planet needs to be considered, see \eg, Eqs. (15) and (16) in the  supplementary material of  \cite{Johansen2015}.   
This is nevertheless only a first order approximation, valid for a relatively low eccentricity. The local frame is not a good approach in case that the eccentricity is not so small.  In contrast, the global approach is conceptually straightforward to conduct  simulations with the planet moving on an eccentric orbit. In this paper, the local method is restricted to planets on circular orbits, while the global method will be applied to  planets on both  circular and eccentric orbits.

\changed{In the local simulation, the pebbles' initial $x$ locations  ($x_0$) range from  $-5 R_{\rm H}$ to  $5 R_{\rm H} $, with an interval  of $10^{-4} R_{\rm H}$, and  $y_0$ is initialized at either $-40R_{\rm H}$ or $40R_{\rm H}$.   The initial radial and azimuthal velocity of the pebble are: 
$ v_{\rm x} =    -2\taus v_{\rm hw}/(1 + \taus^2 )$ and $ v_{\rm y} = 1.5 x_0- v_{\rm hw} /(1 + \taus^2)$ \citep{Weidenschilling1977a} where $\taus \equiv t_{\rm s}\Omega_{\rm p}$ is the dimensionless stopping time (Stokes number). The simulation terminates when the pebble leaves the domain of the shearing box ($|y| >1.05 y_0$), or when the separation between the pebble and the planet is smaller than the planet radius $R_{\rm p}$, indicating a collision.  }
 
In the global simulation, the planet is initiated on a circular ($\ep=0$) orbit (\se{compare}) or on an eccentric orbit ($\ep>0$) with a random true anomaly $\theta$ (\se{ecc}). \changed{ For circular  cases,  $N_{\rm tot}$ pebbles are uniformly distributed along a ring exterior to the planet's orbit at an initial distance $r_{0} = a_{\rm p}(1 + e) + 5 R_{\rm H}$.  However, the orbital phase of the planet affects the accretion efficiency when the planet is on an eccentric orbit.  In such a situation ($\ep>0$),  pebbles are not only distributed azimuthally, but also radially,   from $r_{0}$ to $r_0 + 2\pi v_{\rm r} /\Omega_{\rm K}$.  There are $\sqrt{N_{\rm tot}}$ grid points in both the vertical and radial directions and $N_{\rm tot}$ in total.}  The initial radial and azimuthal velocity of the pebble are: $ v_{\rm r} =    -2\taus v_{\rm hw}/(1 + \taus^2 )$ and $ v_{\rm \phi} = v_{\rm K}  -   v_{\rm hw} /(1 + \taus^2)$.  The simulation terminates when the pebble drifts interior to the orbit of the planet where  $r < a_{\rm p}(1 - e) - R_{\rm H}$, or collides with the planet where $|r - \rp |< R_{\rm p}$.  We set $a_{\rm p} = 1 \AU$ for all simulations in this paper.

For both methods, the equations are numerically integrated with a Runge-Kutta-Fehlberg variable timestep scheme \citep{Fehlberg1969}.  The relative error tolerance is adopted to be $10^{-8}$ to ensure the numerical convergence.


  

%% file: results.tex
\begin{figure*}[tbh!]
    \sidecaption    
      \includegraphics[scale=0.68, angle=0]{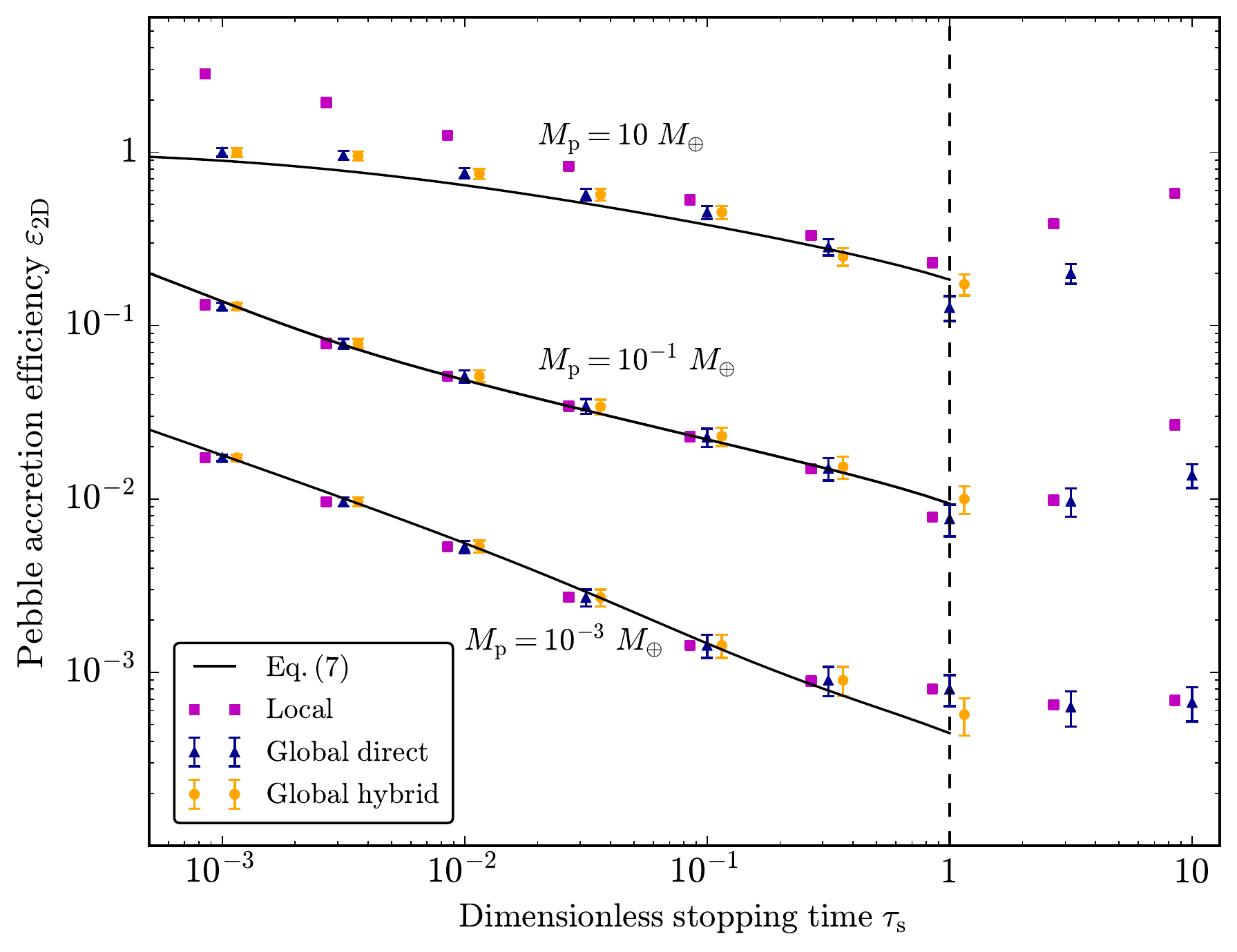}
      \caption{ Pebble accretion efficiency $\varepsilon_{\rm 2D}$ vs $\taus$ and $M_{\rm p}$. Three planet masses ($M_{\rm p} = 10^{-3} \Me$, $10^{-1} \Me$ and $10 \Me$) and nine dimensionless stopping times ($\taus$ ranging from $10^{-3}$ to $10$) are considered. Three different simulations are conducted for each case, the local method (magenta square), global direct method (blue triangle) and global hybrid method (orange circle), respectively.  The errorbar represents the Poission error. Different total number of pebbles are used for different planet mass cases, $N_{\rm tot} = 3\times 10^{4}$ for $\Mp = 10^{-3} \Me$, $N_{\rm tot} = 3000$ for $\Mp = 10^{-1} \Me$, and $N_{\rm tot} = 300$ for $\Mp = 10 \Me$, respectively.   
      The solid line corresponds to the analytical fit described in \se{expression}.  The vertical dashed line gives the boundary where our fit can be applied.   For the global approach, the hybrid method matches the direct method  well  when $\taus \lesssim 0.1$. Local simulations are consistent with  global simulations in the settling regime ($ \taus < 1$) except when the planet is very massive ($\eg, M_{\rm p} =10 \Me$).        }
\label{fig:diff}
\end{figure*}

\section{Results}
\label{sec:results}

Results comparing the different methods for planets on circular orbits  are presented in \se{compare}, and global simulations for planets on eccentric orbits are conducted in \se{ecc}. The analytical fit is given in \se{expression}.  

\subsection{Comparison between different approaches} 
\label{sec:compare}

We conduct simulations with the three different methods (\se{integration}) for planets on circular orbits ($\ep=0$) and compare the obtained pebble accretion efficiencies. We assume that the planet density  is $3 \ \rm gcm^{-3}$ and central star has $1 \Ms$.  
Three planet masses  with varying $\taus$ have been explored. \Fg{diff} shows the pebble accretion efficiency as functions of $\taus$ and $M_{\rm p}$. The symbols and colors depict results from three different methods, and black lines represent our analytical fit expression (to be discussed in \se{expression}). 
For the global method,  the dot and the errorbar represent the mean value from the simulations and the Poisson counting error ($\Delta \varepsilon = \sqrt{N_{\rm hit}}/N_{\rm tot}$), respectively.   Hereafter the efficiency refers to its mean value from the simulations.  We focus on the pebble accretion regime (settling regime)  when  $10^{-3} \lesssim \taus < 1$. But for the additional method comparison, the cases of $\taus \gtrsim 1$ have also been investigated with the local and the global direct methods, in which the gravity of the central star also becomes important during the pebble-planet encounter (referred to as the three-body regime in \cite{Ormel2010}).

The strong coupling approximation (SCA) works well for the small $\taus$ pebble where the gas-pebble coupling time is short. We do not apply the global hybrid method for $\taus >1$ cases. It can be seen in \fg{diff} that the hybrid method (orange circle) well matches with the direct method (blue triangle) when $\taus \lesssim 0.1$.  Moreover, the hybrid algorithm significantly reduces the computational time. 
For a $0.1 \Me$ planet accreting pebbles of $\taus =10^{-2}$, the hybrid integration is $40$ times faster than  the direct integration.  It becomes  $200$ times faster when $\taus$ is  $10^{-3}$.

We also see in \fg{diff} that the results from the local method (red square) and the global methods are in good agreement with each other for the low and intermediate mass planet ($10^{-3} \Me$ and  $0.1 \Me$). Only for the massive planet case ($10 \Me$) the local efficiencies are clearly higher than the global ones. 

There are two reasons for the difference in $\varepsilon_\mathrm{2D}$ between the two methods in the high mass planet case.
First,  there is a risk of multiple counting accreting pebbles for the local method. 
In general pebbles can be accreted by the planet through two ways: (i) immediate accretion upon the first penetration of the planet's Hill sphere; and (ii) secondary accretion after a synodical time of the first close-encounter where $t_{\rm syn} \sim  \Omega_{\rm p}^{-1} r_{\rm p }/\Delta r_{\rm co}$ and  $\Delta r_{\rm co}$ is the width of the co-orbital region of the planet.
In the local frame, we initially place pebbles at both sides of the planet (\ses{local}{integration}). The problem with this setup, however, is that it may amount to simulate the same pebble twice. The risk of double counting occurs in particular when the radial drift of the pebble within a synodical time  is  small compared to the accretion radius, $v_{\rm r} t_{\rm syn} < b_{\rm set}$. 
 It follows that the pebble accretion efficiency of the planet whose mass is higher than  a few Earth masses is overestimated due to this multiple counting effect.      

The other reason is due to the linearization of the equations of motions. In the local method, we ignore the high order terms of $(\Delta r_{\rm p}/\rp)^n$ ($n \geq2$) in the pebble's equation of motion by virtue of the approximation that $\Delta r_{\rm p} \ll  \rp$, where $\Delta r_{\rm p}$ is the distance between the planet and the pebble. The integration domain (the maximum $\Delta r_{\rm p}$ we consider in the local frame) is larger for a more massive planet. Thus, the condition  $\Delta r_{\rm p} \ll  \rp$ breaks down and the first order linear approximation becomes inappropriate for very massive planets.  To conclude, both effects are biased towards high planet mass, and therefore the local results overestimate $\varepsilon$ in the high mass planet case.

We also see in \fg{diff} that the local and the global simulations are inconsistent when $\taus$ becomes larger than $1$.
The two methods give very different results and this mismatch increases with the planet mass.  The pebble's eccentricity can be excited by the planet during the encounter. For pebbles with large stopping time, their eccentricity damping  from the gas drag is much longer than the encounter time ($\simeq$ orbital time). These pebbles should remain at moderate eccentricities during pebble-planet interactions.   However, in the local method pebbles' are initialized on unperturbed ($\ep=0$) trajectories, which is not valid when $\taus$ is large. On the other hand, the global method does not suffer from this effect, since it integrates these pebbles for many complete orbits. These differences become more extreme for larger stopping times and higher planet masses. For the $\Mp = 10 \Me $ and $\taus =10$ case, $\varepsilon_{\rm 2D}$ evaluates to zero in the global simulation, due to the fact that  the pebbles get trapped into resonance with the planet. Of course, resonance trapping cannot be incorporated in a local framework.

In addition, in \fg{diff} the analytical fit expression derived in \se{expression} matches well with both methods  when the pebble accretion is in the settling regime ($10^{-3} \lesssim \taus < 1$). We find that the efficiency increases with the mass of the planet and decreases with the size of the accreted pebbles. The efficiency is increased, because massive planets have a larger impact cross section \eq{reff1}, while small pebbles, because of their slow radial drift, are more likely to encounter the planet.

In summary, the results of the global methods and the local method are mostly in good agreement with each other in the settling regime.  Although computationally more expensive, the global methods are intuitive, and more accurate for higher planet mass ($\Mp $ large than a few Earth mass) and large pebbles ($\taus \gtrsim 1 $).  The global method is arguably the better approach to use, in particular for configurations such as resonance trapping, and, more generally, for eccentric orbits. In the following subsection, we only use the global method to calculate the accretion efficiency.

\subsection{Eccentric pebble accretion}
\label{sec:ecc}
In \se{default}, we describe  the default run for eccentric pebble accretion. In \se{para}, a parameters study is conducted to explore the dependence of pebble accretion efficiency on  the  planet mass ($\Mp$), the stellar mass ($M_{\star}$),  the headwind velocity ($\vhw$) and the dimensionless stopping time ($\taus$) with  respect to the planet eccentricity. 

\begin{figure}[tb]
    \includegraphics[scale=0.5, angle=0]{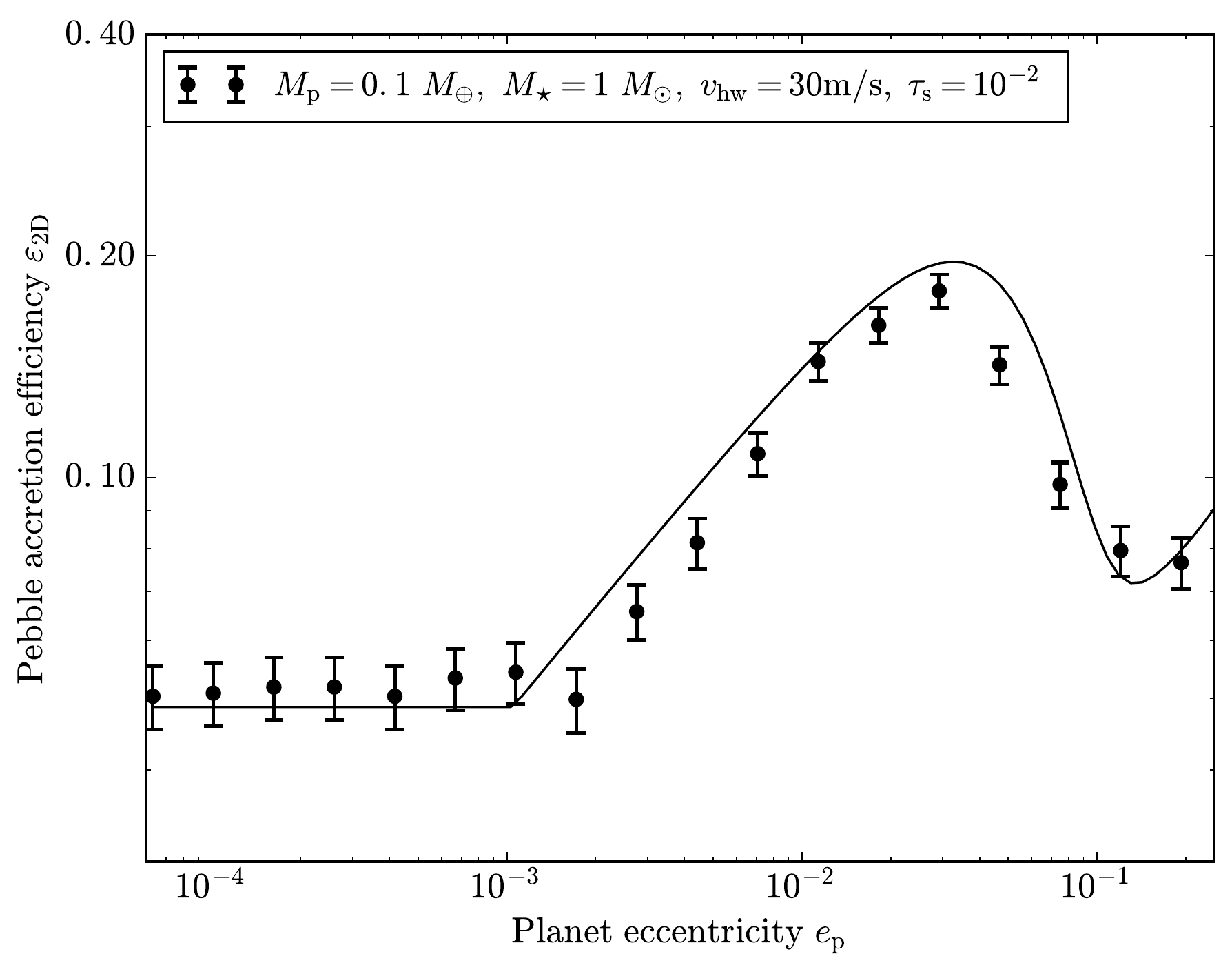}
    \caption{Pebble accretion efficiency vs planet eccentricity for the default run: $\Mp = 0.1 \Me$, $M_{\star} = 1 \Ms $, $ \vhw = 30 \rm \ m \ s^{-1}$, and $\taus = 10^{-2}$.  Eccentricities have been explored from $\ep=6\times 10^{-5}$ to $0.2$. The dot gives the mean value, and the errorbar indicates the Poisson counting error. The solid line represents the analytical fit.  The efficiency peaks around an eccentricity of $0.03$, and  this peak efficiency is four times of the circular efficiency when $\ep \simeq 0$. } 
\label{fig:default}
\end{figure} 

\subsubsection{Default run}
\label{sec:default}
The parameters of  the default model are: $\Mp = 0.1 \Me $, $M_{\star} = 1 \Ms $, $ \vhw = 30 \rm \ m \ s^{-1}$ and $\taus = 10^{-2}$. 
\Fg{default} shows that the pebble accretion efficiency varies with the planet eccentricity.  We conduct simulations with $18$ different eccentricities, logarithmically spaced between $6\times 10^{-5}$ and $0.2$.
For each eccentricity, we perform ten simulations randomly varying with different initial planet phase angle $\theta$.  In this case we set the total number of pebbles $N_{\rm tot}=200$,  $N_{\rm hit}$ is counted from the simulations , and the pebble accretion efficiency is  $N_{ \rm hit}/N_{\rm tot}$.  The black line is the analytical fit, which will be described in \se{expression}.



 
We find that the pebble accretion efficiency varies with eccentricity. When the planet is on a nearly circular orbit ($\ep \lesssim 10^{-3}$), it remains at $0.05$. When the eccentricity keeps increasing, the efficiency first increases and then decreases. It attains a maximum value of  $\varepsilon_{\rm 2D} \simeq 0.18$ when the eccentricity approaches $0.03$. After that, the efficiency quickly reduces to $0.07$ when $\ep \simeq 0.1$.  Clearly, the planet's eccentricity plays an important role in determining the pebble accretion efficiency. At its peak, the efficiency of an eccentric orbit planet ($\varepsilon_{\rm 2D} (\ep)$) is higher than a circular orbit planet by a factor of $4$.

\changed{
By balancing the settling time and the encounter time, it can be shown that the pebble accretion radius is \citep{Ormel2010,Lambrechts2012}  
\begin{equation} 
   b_{\rm set} \sim \sqrt{\frac{GM_p t_s}{\Delta v}},
   \label{eq:reff}
\end{equation} 
 where $\Delta v$ is the relative velocity between the planet and the pebble
 (see  \se{appendix})}.
 
\changed{When a planet is on a circular orbit, the relative velocity  between the pebble and the planet  is either dominated by Keplerian shear or the headwind velocity. 
When a planet is on an eccentric orbit, the relative velocity in addition includes an eccentricity contribution due to the elliptic motion of the planet, which increases with the planet eccentricity. 
Therefore, when $\ep$ is tiny,  $\Delta v$ is very close to the circular case and $\varepsilon_{\rm 2D}$ still remains constant.
The rise of $\varepsilon_{\rm 2D}$  is due to an increase in $\Delta v$ when the eccentric velocity $\sim$$ev_K$ starts to dominate the relative velocity. The flux of pebbles increases. Physically, in the 2D limit,  a planet on a more eccentric orbit sweeps up more pebbles due to a larger pebble feeding zone, resulting in a higher efficiency.  }

The rapid drop of $\varepsilon_{2D}$ at higher $\ep \simeq 0.04$, however, indicates the settling interactions fail. 
\changed{The reason is that  the time for the planet-pebble encounter decreases as  $\Delta v$ (eccentricity) increases.  When the planet-pebble encounter is short, the change of the pebble's velocity due to the gas drag is modest. }
Therefore, the accretion transitions from the settling regime to the ballistic regime, where  the gas drag effect is negligible and the accretion radius reduces to the planet's physical radius (gravitational focusing is unimportant at these high eccentricities). Thus, the accretion efficiency declines significantly during this transition.   

\begin{figure*}[tbh!]
\includegraphics[scale=0.48, angle=0]{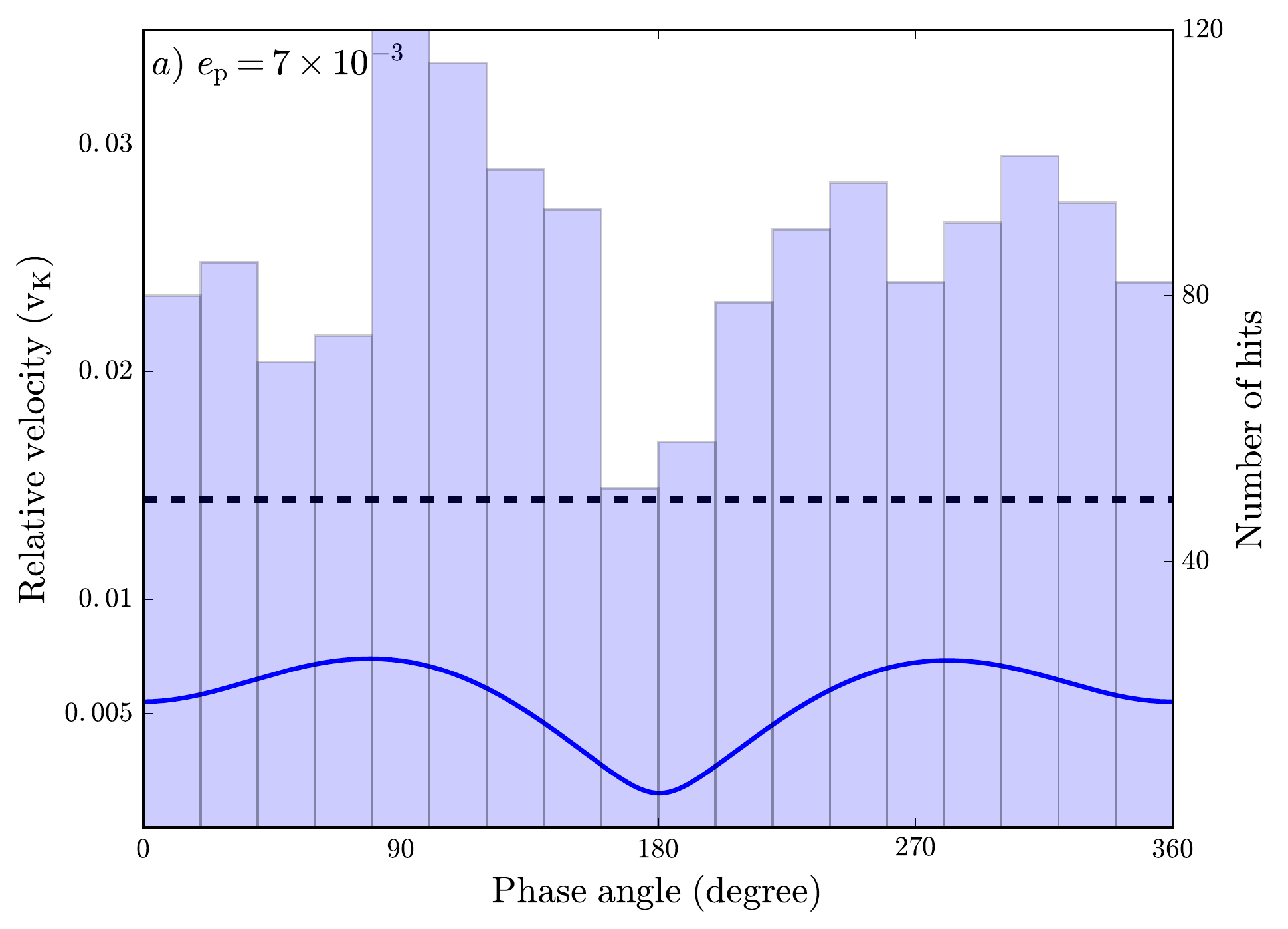}
\includegraphics[scale=0.48, angle=0]{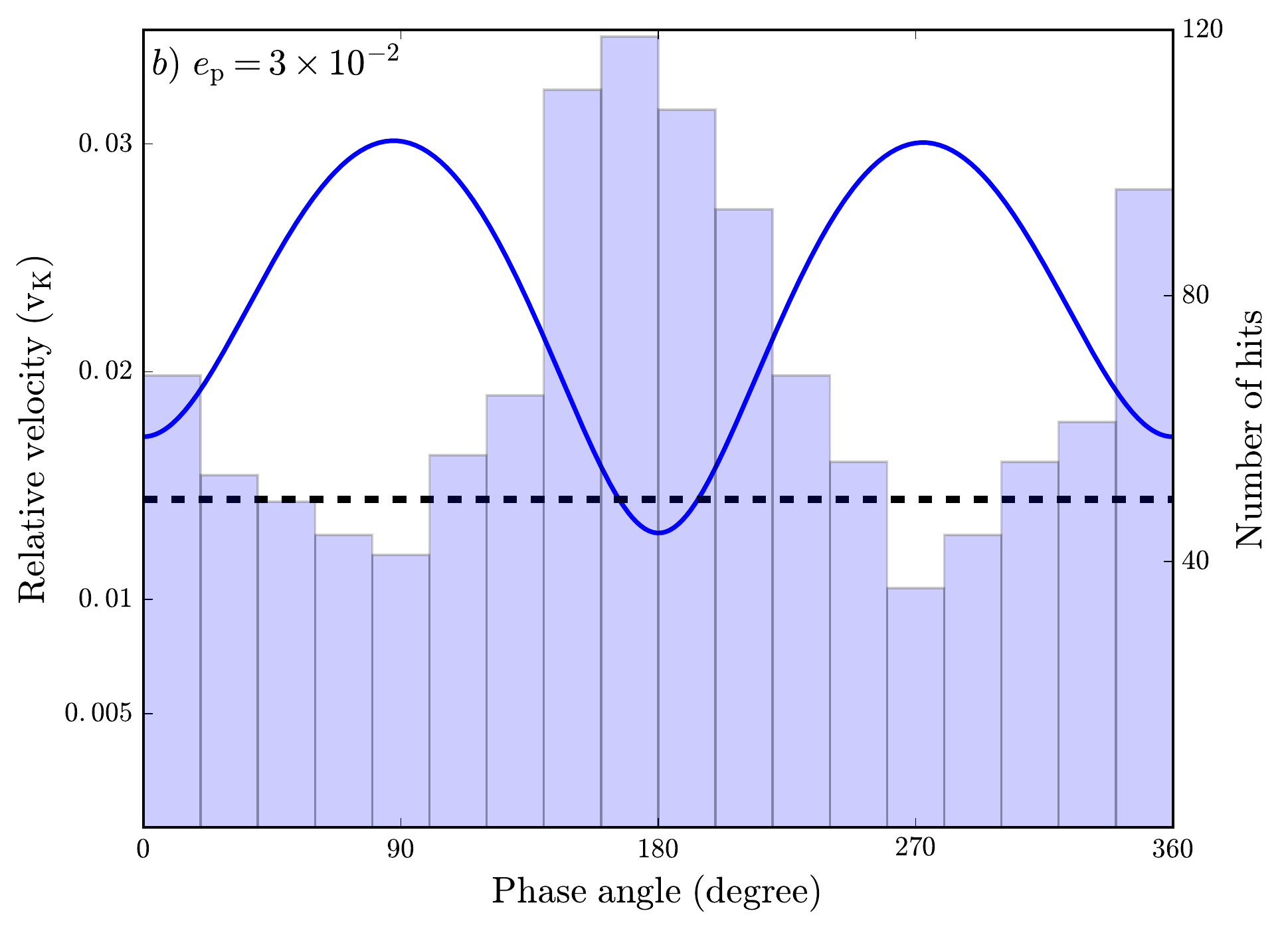}
\caption{\changed{ Relative velocity between the planet and pebbles (dark blue line) and number of accreted pebbles (light blue bars) as function of the planet's orbital phase angle. Perihelion is at $0 ^\circ$ and aphelion is at $180 ^\circ$. The planet eccentricity is  $\ep = 0.007$ in the left and  $\ep = 0.03$ in the right panel, respectively.  Other parameters are the same for the two panels:  $\Mp = 0.01 \Me $, $M_{\star} = 1 \Ms $, $\taus = 10^{-2}$, $v_{\rm hw} = 60 \ \rm m/s$ and $N_{\rm tot} =300\times300$.  The black dashed line represents the transition velocity from the settling to the ballistic accretion, $v=v_\ast$ (\eq{vstar}).}
} 
\label{fig:angle}
\end{figure*} 

\changed{In order to analyse this issue in detail,  we investigate at which location (planet's orbital phase angle) pebbles are captured by the planet. The capture condition is defined as the separation between the pebble and the planet is smaller than the accretion radius  ($ |r-r_{\rm p} |< b_{\rm set}$).   In order to clarify the effect we focus on the low mass planet when it is initially in the headwind regime,  therefore we adopt: $\Mp = 0.01 \Me $, $M_{\star} = 1 \Ms $, $\taus = 10^{-2}$, $\vhw = 60 \ \rm m/s$.  We artificially set the planet radius $10$ times smaller than its normal value to eliminate any ballistic accretion. 
\Fg{angle} plots the relative velocity between the planet and pebble (left y axis) and the number of accreted pebbles (right y axis) as  functions of the planet's phase angle for $\ep = 7 \times 10^{-3}$ (\fg{angle}a)  and $\ep = 3 \times 10^{-2} $ (\fg{angle}b).
The radial and azimuthal velocity of a planet on eccentric orbit are $v_{\rm K} (\rp) \ep \sin\theta /\sqrt{1-\ep^2} $ and $v_{\rm K} (\rp)(1 + \ep \cos\theta) /\sqrt{1-\ep^2} $, where $\theta$ is the true anomaly of the planet \citep{Murray1999}.  Assuming that the orbit of the pebble is not affected by the planet, its radial and azimuthal velocity are $-2 \vhw \taus v_{\rm K}/(1+ \taus^2) $ and $v_{\rm K}- \vhw/(1 + \taus^2)$. Therefore, the relative velocity between the planet and the pebble can be analytically calculated, which is shown by the blue lines of \fg{angle}.
}

\changed{We find that in \fg{angle} the number of accreted pebbles correlates with $\Delta v$. In the low eccentricity case (\fg{angle}a), more pebbles are accreted when $\Delta v$ is higher. For instance, when the pebble moves on a sub-Keplerian orbit it has the lowest relative velocity with the planet at aphelion ($180 ^\circ$).  We find that at this location the planet accretes the least number of pebbles. 
Since in \eq{reff} $b_{\rm set} \propto \Delta v^{-1/2}$,  the accretion radius is largest at aphelion. However, the accretion rate correlates with $b_{\rm set} \Delta v$ in the 2D limit and with $b_{\rm set}^2 \Delta v$ in the 3D limit \citep{Johansen2015,Morbidelli2015}.  Therefore, $\varepsilon_{\rm 2D}$ increases as $\Delta v^{1/2}$. Less pebbles are accreted at aphelion than elsewhere. 
}

\changed{ On the other hand, when the eccentricity is high, the pebble accretion rate displays the opposite dependence on the phase angle (\fg{angle}b), i.e., more pebbles are accreted  at lowest $\Delta v$ (aphelion).  This is because in this high eccentricity case, $\Delta v$ becomes too high for the settling.  Thus, only pebbles with relatively low $\Delta v $ are able to be accreted by the planet. The transition velocity from the settling to the ballistic accretion is marked by the black dashed line in \fg{angle} (\eq{vstar}). }

\subsubsection{Parameter study}
\label{sec:para}
To investigate the effects of different parameters on the pebble accretion efficiency, we conduct a  parameter study by varying the planet mass ($\Mp$), the stellar mass ($M_{\star}$), the headwind velocity ($\vhw$) and the dimensionless stopping time ($\taus$). We explore four planet masses ($10^{-3}\Me, 10^{-2}\Me, 10^{-1}\Me, 1\Me $), three stellar mass ($3\Ms, 1\Ms, 0.3\Ms $), three headwind velocities ($15\ \rm m \ s^{-1}, 30\ \rm m \ s^{-1}, 60\ \rm m \ s^{-1}$) and three dimensionless stopping times ($10^{-3}, 10^{-2},10^{-1}$).  Nine of these simulations are illustrated in \Tb{tab1} and \fg{para}.

\subsubsection*{Planet mass}
\fg{para}a shows the efficiency as a function of the eccentricity for three different planet masses. The other  parameters are the same as the default values.  The default case ($\Mp= 0.1 \Me$), the massive planet case ($\Mp= 1 \Me$), and the less massive planet case ($\Mp= 0.01\Me$) are shown in black, dark red and light red, respectively.

The three cases exhibit a similar trend with eccentricity.  The efficiency curves are initially flat, gradually increase with $\ep$ later and drop to lower values. For the massive planet case, the efficiency is $ 0.22$ when the planet is on a circular orbit.  The maximum efficiency attains $0.59$ when $\ep=0.05$. For the less massive planet case, the efficiency of a circular orbit planet is $0.013$  and the maximum efficiency is $0.04$ when $\ep$ is close to $0.01$.  Comparing the three cases, we find that (i) the efficiency increases with the planet mass for all eccentricities; more massive planets accrete pebbles more efficiently (the same as the circular case in \se{compare}). (ii) The transition eccentricity where the efficiency attains its maximum increases with the planet mass. This is because the transition occurs when the relative velocity $\Delta v$ is so high that the settling accretion enters the ballistic accretion (see \se{appendixset}).    Since a more massive planet has a stronger gravitational attraction that enables to accrete more eccentric pebbles, the transition velocity increases with the planet mass (\eq{vstar}). Consequently, the  transition eccentricity also increases with the planet mass.

\subsubsection*{Stellar mass}
\fg{para}b illustrates the dependence of the stellar mass on the efficiency where the rest of parameters are identical as the  default case.  The default case ($M_{\star} =1 \Ms$) is shown in black, and the massive giant star case ($M_{\star}= 3 \Ms$) and the low-mass M dwarf star case ($M_{\star}= 0.3 \Ms$) are given in light purple and dark purple, respectively.

For the massive star, the efficiency is $ 0.026$ when the planet is on a circular orbit.  The maximum efficiency attains $0.09$ when $\ep=0.03$. For the less massive star, the efficiency for a circular orbit planet is $0.1$  and the maximum efficiency is $0.34$ when $\ep = 0.04$.  Comparing the three cases, we find that (i) the efficiency decreases with the stellar mass.  (ii) The transition eccentricity decreases with the stellar mass. Both effects can be explained from the fact that  a planet around a less massive star has a larger Hill sphere and therefore can accrete more pebbles.  

\begin{figure*}[tbh!]
      \includegraphics[scale=0.5, angle=0]{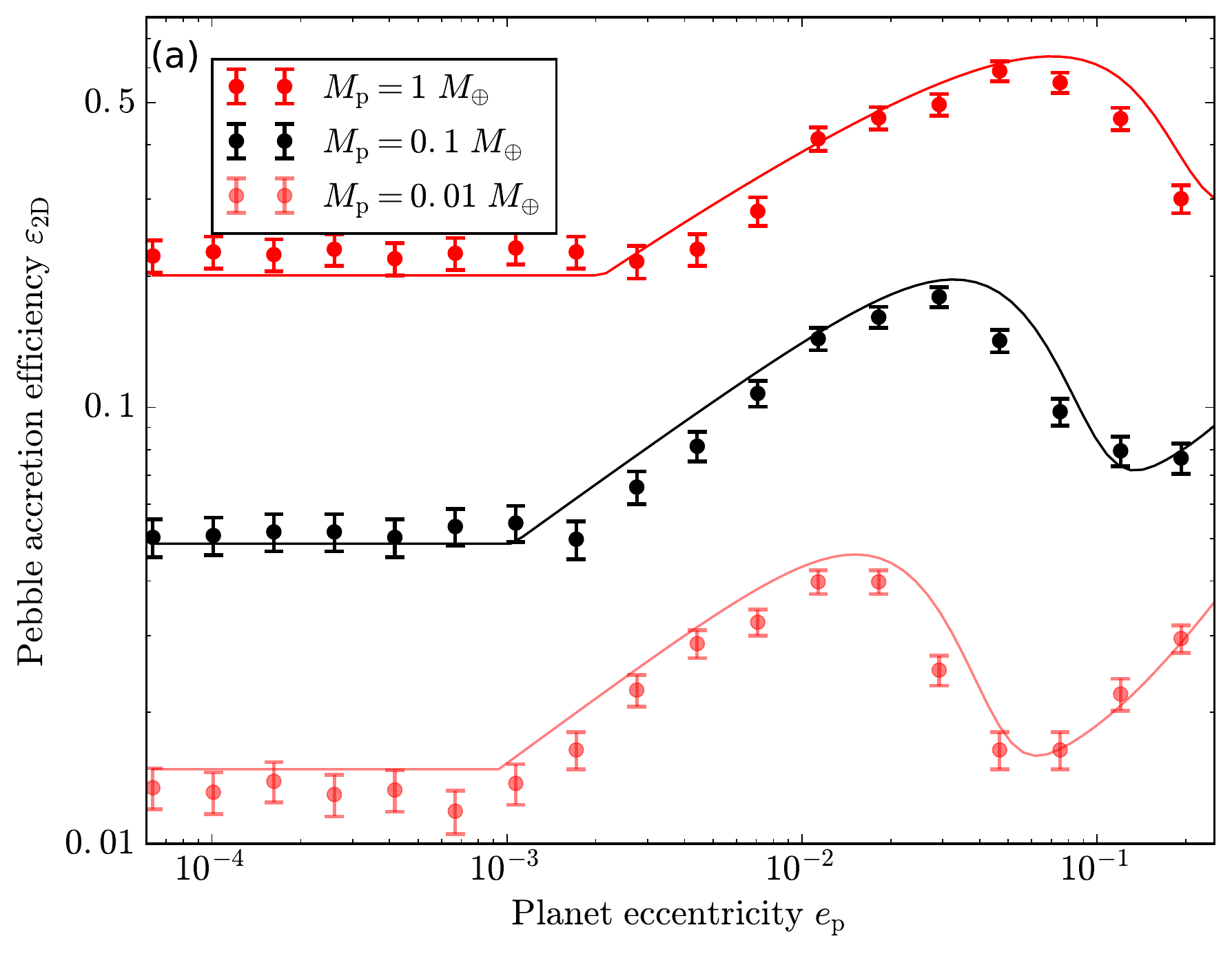}
       \includegraphics[scale=0.5, angle=0]{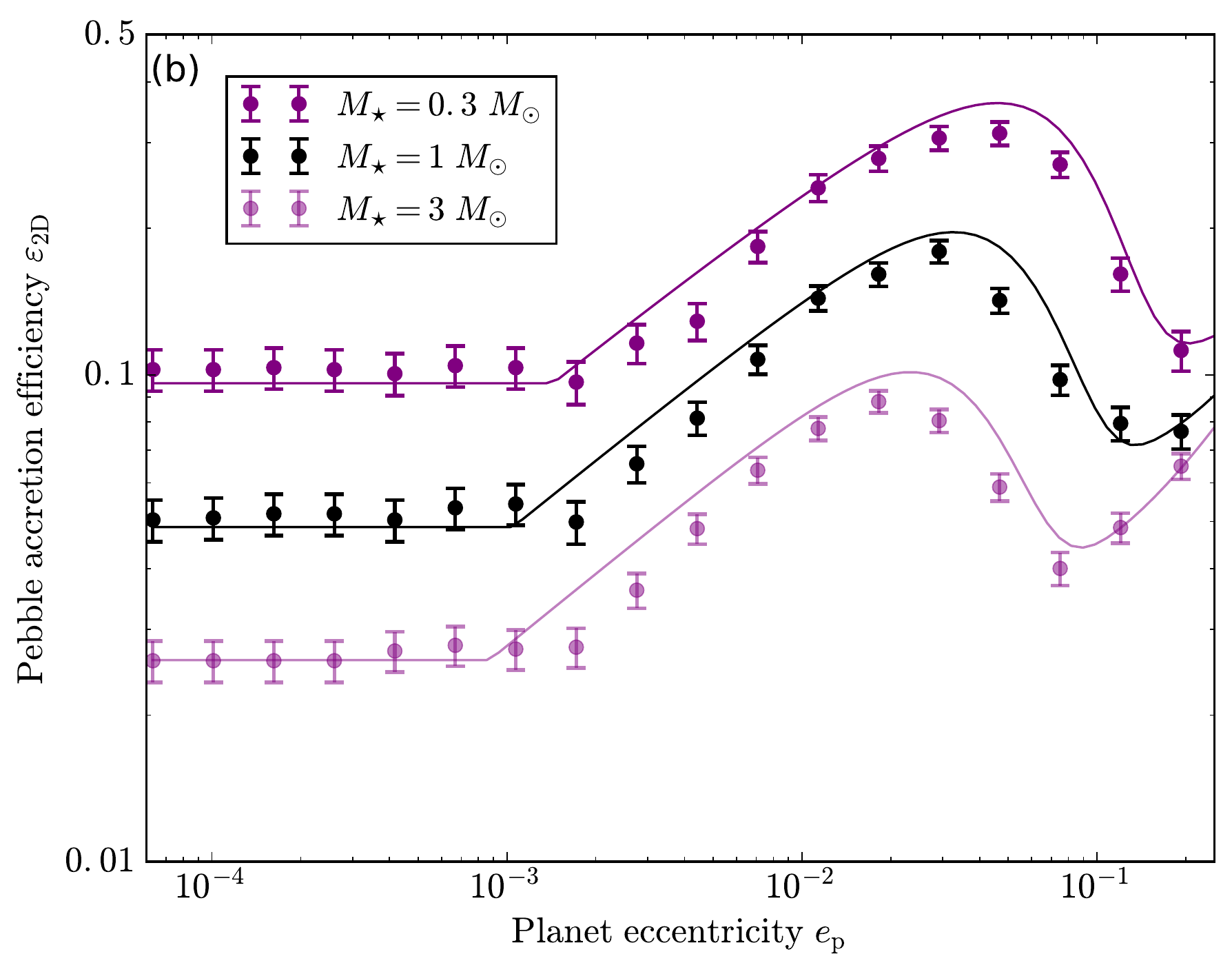}
       \includegraphics[scale=0.5, angle=0]{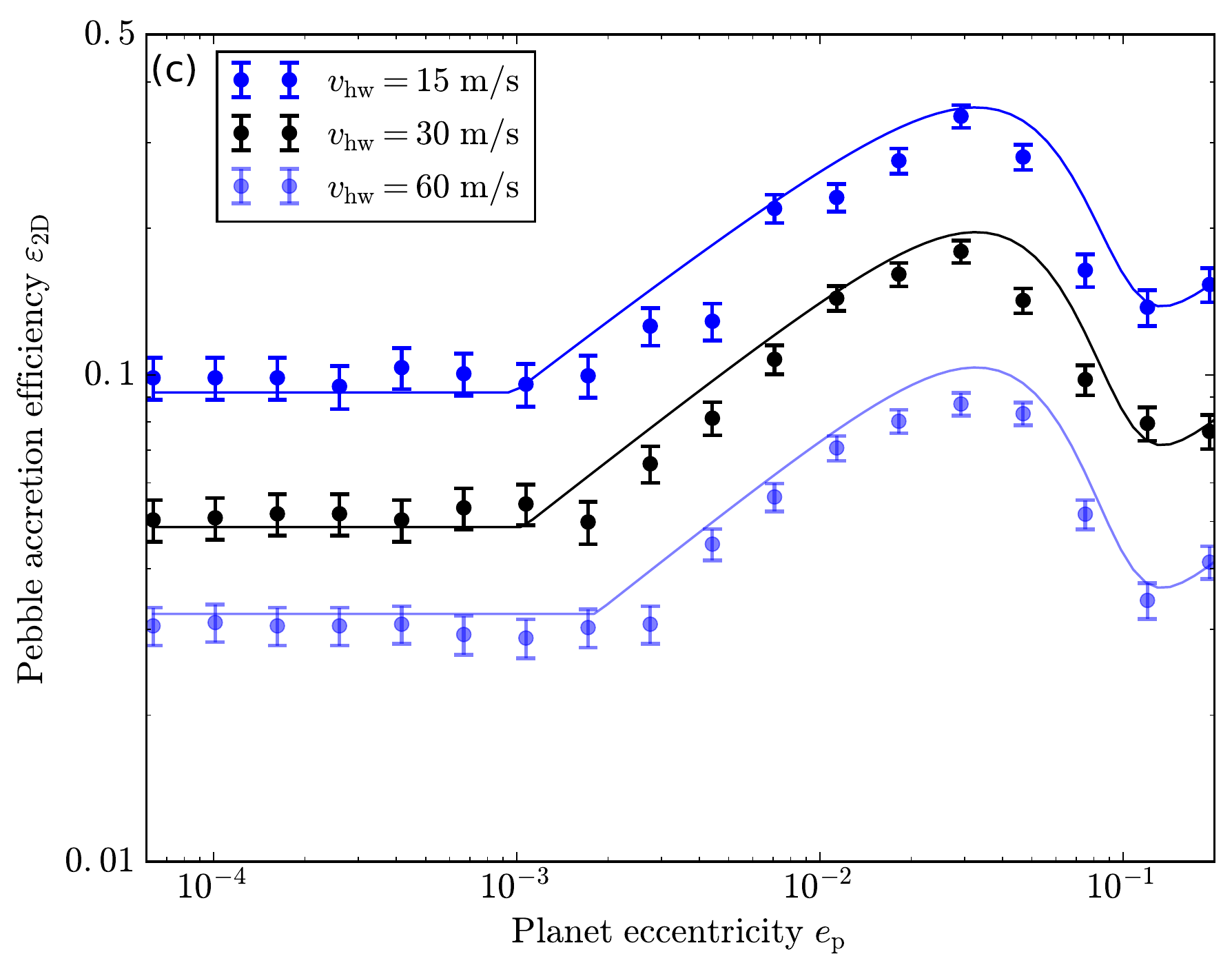}
        \includegraphics[scale=0.5, angle=0]{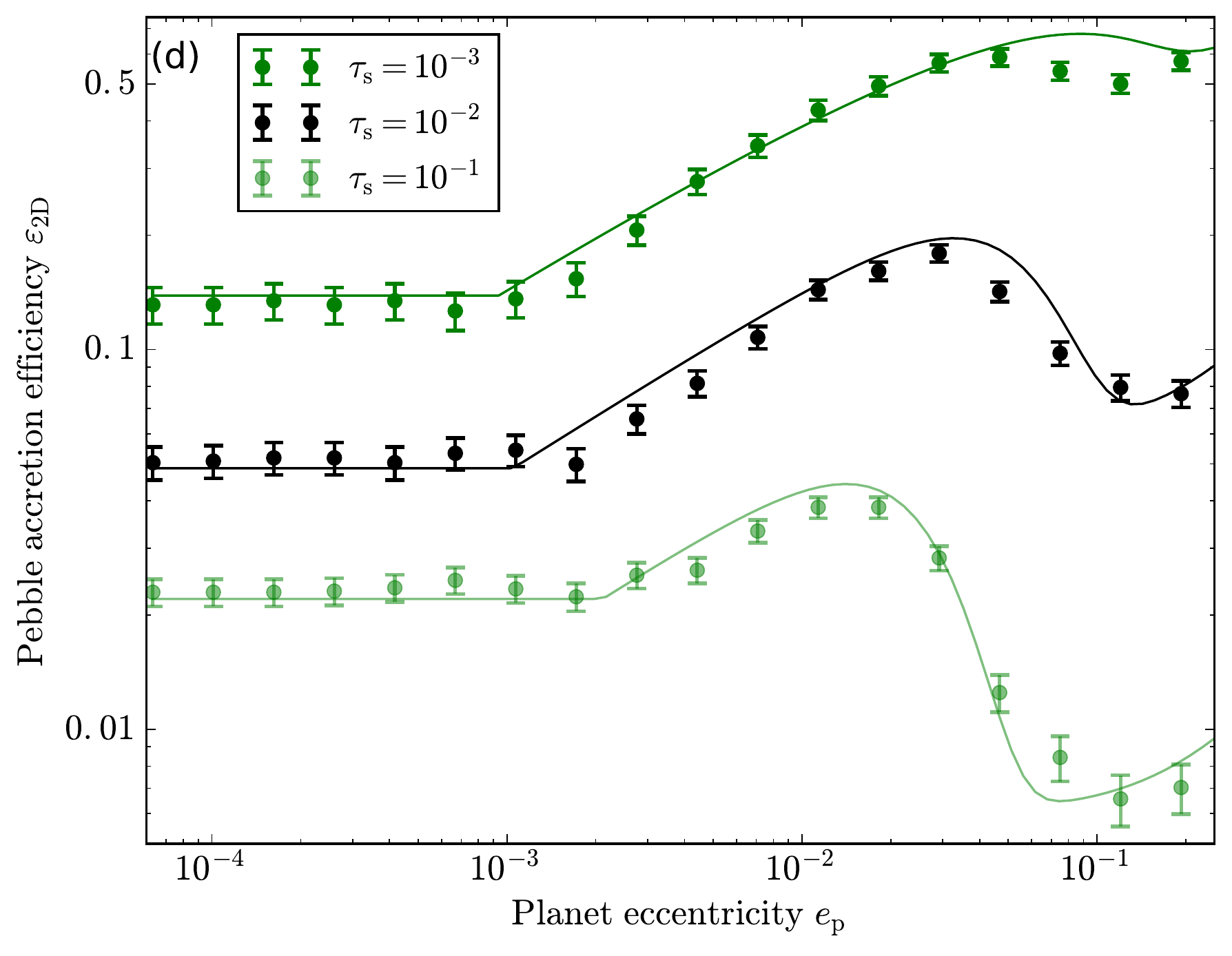}
\caption{ Pebble accretion efficiency vs  planet eccentricity. The default run is shown in black, with $\Mp = 0.1 \Me $, $M_{\star} = 1\Ms$, $\vhw = 30 \rm \ m \ s^{-1}$ and $\taus = 10^{-2}$. In each panel one parameter varies and the other three keep the same.  The upper left panel: three different planet masses,  $\Mp = 1 \Me$, $\Mp = 0.1 \Me$ and $\Mp = 0.01 \Me$; The upper right panel: three different stellar masses,  $M_{\star} = 3 \Ms$, $ M_{\star} = 1 \Ms$ and $M_{\star} = 0.3 \Ms$; The lower left: three different headwind velocities,  $\vhw = 15 \rm \ m \ s^{-1}$, $\vhw = 30 \rm \ m \ s^{-1}$ and $\vhw = 60 \rm \ m \ s^{-1}$; The lower right: three dimensionless stopping times,  $\taus = 10^{-3} $, $\taus=10^{-2}$ and $\taus=10^{-1}$.  The dot gives the mean value, and the errorbar indicates the Poisson counting error. The solid line is the analytical fit of \eq{efficiency}.  The efficiency increases with $M_{\rm p }$, and decreases with $M_{\star}$, $\vhw$ and $\taus$.} 
\label{fig:para}
\end{figure*}

\subsubsection*{Headwind velocity}
\fg{para}c  plots three cases with different headwind velocities. The disk headwind now varies with low speed ($\vhw= 15 \ \rm m \ s^{-1}$) and high  speed ($\vhw= 60 \ \rm m \ s^{-1}$)  shown in dark blue and light blue, respectively. Efficiencies obtained from the default run ($\vhw =30 \ \rm m \ s^{-1}$) are shown in black.

For the low headwind case, the efficiency of pebble accretion is $0.10$ when the planet is on a circular orbit and the maximum efficiency is $0.33$. For higher headwind velocity case, the efficiency when the planet is on a circular orbit is $0.03$,  and the maximum efficiency is $0.08$.  
We conclude that (i) the efficiency decreases with the headwind velocity.  Because  the headwind is related with the radial drift velocity of the pebble, a lower headwind speed results in a slower radial drift, and the planet is therefore able to accrete more pebbles. Furthermore, (ii) the eccentricity for which the maximum efficiency is attained, is independent of the headwind velocity. This is because as the eccentricity increase, $\Delta v$ is dominated by the eccentricity velocity but not  the headwind velocity. Therefore, the transition velocity between the settling and the ballistic regime is independent of  the headwind velocity (\eq{vstar}).
\subsubsection*{Dimensionless stopping time}
Finally, three cases with different $\taus$ are shown in \fg{para}d.  The default case ($\taus =10^{-2}$) is shown in black while small pebbles  ($\taus =10^{-3}$) and large pebbles ($\taus =10^{-1}$) are shown in dark green and light green, respectively. 

For  $\taus= 10^{-3}$ the efficiency of pebble accretion is $ 0.13$ when the planet is on a circular orbit. When $\ep \simeq 0.06$, the efficiency attains a maximum value of $0.60$. For $\taus= 10^{-1}$ the efficiency of a circular orbit planet is $0.02$.  The maximum efficiency is close to $0.04$ when $\ep= 0.01$. 
It is clear that the planet is more efficient at accreting small size pebbles due to their slow radial drift, which is consistent with the results of circular pebble accretion in \se{compare}.

In addition, we find that the ratio of the maximum $\varepsilon_{\rm 2D}$ to  $\varepsilon_{\rm 2D}$ when $\ep \simeq 0$ is high for $\tau_s=10^{-3}$ and low for $\tau_s=10^{-1}$. The reason is that the settling interactions extend to higher $\ep$ when $\taus$ is small. Therefore, the transition to the ballistic regime occurs at higher $\ep$ and the peak efficiency has a stronger boost for small $\taus$.

To summarize, the pebble accretion efficiency $\varepsilon_{\rm 2D}$ is an increasing function of the planet-to-star mass ratio ($\qp = \Mp/M_{\star}$),  a decreasing function of the headwind velocity ($\vhw$) and the dimensionless stopping time of the pebble ($\taus$).  The efficiency is  independent of eccentricity when it is relatively small ($\ep \lesssim 10^{-3}$), gradually increases with eccentricity when it is moderate ($\ep \simeq 10^{-2}$),  and drops quickly when the eccentricity becomes relatively high ($\ep \gtrsim 0.1$).

\begin{table*}
    \centering
    \caption{Model set-up in a parameter study (planet density is $ 3 \ \rm gcm^{-3}$)}
    \begin{tabular}{lllllllllll|}
        \hline
        \hline
        Run    & $M_{\rm p }$ &$M_{\star }$  & $v_{\rm hw}$  & $\taus$  & $N_{\rm tot}$ \\
                &   ($ \rm M_{\oplus}$)   & ($ \Ms$)&  ($\rm m \ s^{-1}$) &  \\
        \hline
    \#1 default & $0.1 $ &$1$ & 30    & $10^{-2}$  & \changed{$45\times45$  }  \\
    \#2 massive planet & $1 $ &$1$  & 30    & $10^{-2}$ & \changed{$25\times 25$}     \\
    \#3 low-mass planet  & $0.01 $ & $1$  & 30    & $10^{-2}$ &  \changed{$80 \times 80$}     \\
     \#4 massive star & $0.1$ &$3$  & 30    & $10^{-2}$ & \changed{$ 64\times 64 $}    \\
    \#5 low-mass star  & $0.1 $ & $0.3$  & 30    & $10^{-2}$ &  \changed{$ 32 \times 32$ }     \\
    \#6  low headwind velocity & $0.1 $ &$1$  & 15    & $10^{-2}$  &\changed{$32\times 32$ }    \\
    \#7  high headwind velocity & $0.1 $ & $1$ & 60    & $10^{-2}$   &\changed{$64\times 64$}  \\
     \#8  small stopping time & $0.1 $ & $1$ & 30    & $10^{-3}$  & \changed{$25\times 25$}   \\
      \#9  large stopping time & $0.1 $ &$1$ & 30   & $10^{-1}$ &   \changed{$ 80\times 80$} \\
          \hline
        \hline
    \end{tabular}
    \label{tab:tab1}
\end{table*}

\subsection{Analytical fit expression}
\label{sec:expression}
\changed{In this subsection, we present analytical fitting formulas, the detailed derivations of these formulas are described at length in the appendix.} 

We find that the obtained accretion efficiency in the settling regime can be well fitted by
\begin{equation}
    \varepsilon_\mathrm{set,2D} 
  = \changed{0.32} \sqrt{\frac{\qp }{\taus \eta^2} \left( \frac{\Delta v}{v_{\rm K}} \right) } f_\mathrm{set},
    \label{eq:eps2d-set}
\end{equation}
where $\qp $ is the mass ratio of the planet to the central star.
 The relative velocity between the planet and the pebble ($\Delta v$) is
\begin{equation}
   \Delta v = {\rm max} \left(v_{\rm cir},v_{\rm ecc} \right),
\end{equation}
Here $v_{\rm cir}$  is the relative velocity between the circular orbit planet and the pebble. We adopt an expression of $v_{\rm cir}$ that combines the headwind and the shear regimes,
\begin{equation} 
   v_{\rm cir} =  \left[1 + \changed{5.7} \left(\frac{\qp}{\qc}\right) \right]^{-1} v_{\rm hw} + v_{\rm sh}, 
   \label{eq:vcir}
\end{equation}
where $v_{\rm hw} = \eta \vk$, $v_{\rm sh} =\changed{0.52} (\qp \taus)^{1/3} \vk$ and the transition mass ratio for two regimes $\qc = \eta^3/\taus$. 
\changed{ The eccentric velocity of the planet relative to its circular Keplerian value is $v_{\rm ecc}$,} where we fit
   \begin{equation}
    v_{\rm ecc} = \changed{0.76} \ep \vk.
       \label{eq:vecc}
\end{equation} 
In the epicycle approximation, it can be shown that the velocity relative to a circular orbit ranges from a minimum of $\ep v_{\rm K}/2$ to maximum of $\ep v_{\rm K}$ \citep[\eg][]{Johansen2015}. Our numerical coefficient of \changed{$0.76$} falls between these limits. Once $\eta>\ep$, $\Delta v$ will be dominated by the headwind, also consistent with the analytical expression by \citep{Johansen2015}.

When $ \Delta v $ becomes higher, the accretion is no longer in the settling regime. We adopt an exponential  function  $f_\mathrm{set}$
 to express the decay of $ \varepsilon_\mathrm{set,2D}$, 
\begin{equation}
    f_\mathrm{set}  = \exp{\left[-0.5 \left( \frac{\Delta v}{v_{\ast}} \right)^{2}\right]},
\label{eq:fset}
\end{equation}
and  
\begin{equation}
v_{\ast} = ( \qp /\taus)^{1/3} v_{\rm K} 
\label{eq:vstar}
\end{equation}
is the transition velocity from the settling to the ballistic regimes \changed{(see derivations in \se{appendixset})}.

The accretion efficiency in the ballistic regime is
\begin{equation} 
 \varepsilon_{\rm bal,2D} 
 =  \frac{ R_{\rm p}} { 2\pi   \taus \eta  r_{\rm p} }\sqrt{ \frac{2 \qp r_{\rm p} }{R_{\rm p}} +  \left( \frac{  \Delta v }{v_{\rm K} } \right)^2   } \left(1 -f_{\rm set}\right).
  \label{eq:effbl0}
 \end{equation}
 It is important to note that $\varepsilon_{\rm set,2D}$ is independent of $r_{\rm p}$ and $R_{\rm p}$, whereas  $\varepsilon_{\rm bal,2D}$ is related with  the ratio of above two quantities, $R_{\rm p}/r_{\rm p}$.

The total accretion efficiency is 
\begin{equation} 
   \varepsilon_{\rm 2D}  = \varepsilon_{\rm set,2D} + \varepsilon_{\rm bal,2D}
\end{equation}

The above recipe for $\varepsilon_{\rm 2D}$ is calculated based on rate  expression. It is therefore not guaranteed that the accretion probability $\varepsilon_\mathrm{2D}$ remains no larger than unity  ($\varepsilon_{\rm 2D}\leqslant 1$). 
The situation is analogous to radioactive decay, where, for a decay rate of $\lambda$, the probability of decaying after a time is $P=1-\exp(-\lambda t)$.
We can therefore correct for this effect, simply by redefining $\tilde{\varepsilon}_{\rm 2D}  = 1 - \exp(-\varepsilon_{\rm 2D})$ 
Clearly, in such a expression, $\tilde{\varepsilon}_{\rm 2D} = \varepsilon_{\rm 2D}$ when  $ \varepsilon_{\rm 2D}\ll1$ and  $\tilde{\varepsilon}_{\rm 2D} \simeq 1$ when  $ \varepsilon_{\rm 2D} \geqslant 1$.
Accounting for these probabilistic nature of $\varepsilon$, we give the efficiency expression as, 
  \begin{equation} 
      \varepsilon_{\rm 2D}  \rightarrow
 1 - \exp(-\varepsilon_{\rm 2D}).
 \label{eq:efficiency}
 \end{equation}
to ensure that $\varepsilon \le 1$.
In \fg{diff} and \fg{para}, we find that this analytical fit expression (\eq{efficiency}) agrees with the simulations quite well for  planets on circular orbits  as well as on eccentric orbits. 

\changed{ \cite{Lambrechts2014} also calculate the efficiency when the planet is on a circular orbit.  In the shear-dominated regime,  \eq{eps2d-set} can be written as 
\begin{equation}
\begin{split}
\varepsilon_{\rm sh}  &=
	 0.1 \left( \frac{M_{\rm p}}{M_{\oplus}} \right)^{2/3} \left( \frac{\eta}{10^{-3}} \right)^{-1} \left( \frac{\tau_{\rm s}}{0.1} \right)^{-1/3} \\
	& \simeq 0.022 \left( \frac{M_{\rm p}}{M_{\oplus}} \right)^{2/3}  \left( \frac{\tau_{\rm s}}{0.1} \right)^{-1/3}  \left( \frac{r}{10 \rm \AU} \right)^{-1/2}
 \end{split}
\label{eq:efficiency2}
\end{equation}
The latter expression of the above formula adopts the same disk model as \cite{Lambrechts2014}.
Comparing \eq{efficiency2} with \cite{Lambrechts2014}'s Eq.(33), we find that  the scaling relations are identical, and our prefactor is  $35 \%$ lower than theirs.}
 
 In the headwind-dominated regime,  \eq{eps2d-set} can be written as 
 \begin{equation}
\begin{split}
\varepsilon_{\rm hw}  &=
	 0.055   \left( \frac{M_{\rm p}}{M_{\oplus}} \right)^{1/2} \left( \frac{\eta}{10^{-3}} \right)^{-1/2} \left( \frac{\tau_{\rm s}}{0.1} \right)^{-1/2}. 
 \end{split}
\label{eq:efficiency_hw}
\end{equation} 
The transition mass from the headwind to the shear regimes is  
 \begin{equation}
 M_{\rm hw/sh} = 0.03  \left( \frac{\eta}{10^{-3}} \right)^{3}  \left( \frac{\tau_{\rm s}}{0.1} \right)^{-1}M_{\oplus}.
\end{equation}

In the eccentricity-dominated regime,  \eq{eps2d-set} can be written as  (neglecting $f_{\rm set}$) 
 \begin{equation}
\begin{split}
\varepsilon_{\rm ecc}  &=
	 0.055   \left( \frac{e_{\rm p}}{10^{-3}} \right)^{1/2}   \left( \frac{M_{\rm p}}{M_{\oplus}} \right)^{1/2} \left( \frac{\eta}{10^{-3}} \right)^{-1}\left( \frac{\tau_{\rm s}}{0.1} \right)^{-1/2}.
 \end{split}
\label{eq:efficiency_e}
\end{equation} 
We find that  in this regime $\varepsilon_{\rm set,2D} $ increases as $e_{\rm p}^{1/2}$.

\subsection{Neglected effects}
In deriving the expression above, we have neglected several effects:
\begin{enumerate}
    \item Evaporation/molten of pebbles for planetesimals travelling on supersonic velocity (relevant for $\ep >\hg$).
        One caveat is that planetesimals on eccentric orbits can produce bow shocks as they move supersonically through the disk gas (\ie, $\ep > \hg$, \cite{Morris2012}). The surrounding gas temperature is raised due to  energy deposition at the shock front. \changed{Therefore, pebbles are likely to be sublimated/melted before they settle onto the planetesimal, depending on the gas density and pressure (\eg, Fig. 6 of \cite{Brouwers2017}), the planetesimal's mass, and the pebble's size and composition \citep{Hood1998,Miura2005}.} The detail treatment is beyond the scope of this paper; here we do not solve for the thermal balance of the pebble. Nevertheless,  stopping the mass growth due to the pebble evaporation is most relevant for low-mass planetesimals.  For high mass planets capable of accreting the primordial disk gas, evaporating pebbles may not reach the surfaces of the cores, but still enrich their envelopes \citep{Venturini2016,Alibert2017,Brouwers2017}.
    \item Aerodynamic deflection \citep{Guillot2014,Johansen2015,Visser2016}, relevant for very low-mass planetesimals and small $\taus$ pebbles.  When the planet mass is very low, its impact radius reduces to the physical radius. 
        However, when the  stopping time of the pebble is short compared to the crossing time of the planetesimal ($\sim R_{\rm p}/v_{\rm hw}$),  the pebble will follow gas streamlines and avoid accretion.This  occurs for typical planetesimal size $R_{\rm p}  \lesssim 100 \ \rm km$ and $\taus \lesssim 10^{-3}$.
    \item Pre-planetary atmosphere formation, relevant for $b_{\rm set} <R_\mathrm{\rm Bondi}$. Once the planet's surface escape velocity becomes larger than the local sound speed, a planet will start to accrete the disk gas, creating a pre-planetary atmosphere.
    The gas density in planetary atmosphere is higher compared to the surrounding disk gas. When the pebble  enters the planetary atmosphere, gas drag becomes larger and the accretion cross section also increases due to this enhanced drag. 
    However, for pebble accretion, the accretion cross section ($b_{\rm set} \simeq \taus^{1/3}R_{\rm H}$ in the shear regime) can be as large as the planet's Hill sphere. The radius of the planetary atmosphere cannot exceed the Bondi radius, $R_{\rm Bondi} = G \Mp/c_{\rm s}^2$, where $c_{\rm s}$ is the sound speed at the planet's location.  As long as $R_{\rm Bondi} < b_{\rm set}$ (or $q_p<2.3\tau_s^{1/2}h_\mathrm{gas}^3$ in dimensionless units), it is therefore appropriate to neglect the atmosphere enhancement on pebble accretion. For a planet accreting $\taus = 10^{-2}$ pebbles in the disk with a typical scale height $\hg = 0.05$, this condition is justified as long as the planet is less than $10 \Me$.  

    In addition, we assume that the gas moves on an unperturbed sub-Keplerian orbit. In reality the planet will perturb the gas flow.  A natural scale on which these effects appear is again the planet's Bondi radius \citep{Ormel2015}. Therefore, the flow pattern effect on pebble accretion is also minor when $R_{\rm Bondi} < b_{\rm set}$.   
    
\item Pebble isolation mass.  When the planet is massive enough to strongly perturb the disk gas, a gap can be formed  in the vicinity of the planet and inverses the locally gas pressure gradient. This process  truncates pebbles' inward drift, and therefore terminates pebble accretion.  The planet mass is defined as  the pebble isolation mass ($M_{\rm p, iso}$). Approximating the gap opening mass mass \citep{Lin1993},   $M_{\rm p, iso}$ is  around $20 \Me$ for a solar mass star ($M_{\rm p, iso} \sim h_{\rm gas}^3 M_{\star}$, \cite{Lambrechts2014b,Bitsch2018}).  In our study, we consider the planet mass in a range of  $10^{-3} \Me \lesssim \Mp \lesssim 10 \Me < M_\mathrm{p,iso}$.
\end{enumerate}

%% file: application.tex
\section{Application: formation of a secondary planet}
\label{sec:application}
In this section we envision an already-formed primary planet (a gas giant) and a planetary embryo for the future second planet. 
We discuss how secondary planet formation is aided by a higher pebble accretion efficiency at resonance locations, where eccentricities are excited by the primary giant planet. 

Based on the observational statistics of the occurrence rate,  $10\%$-$20\%$ of solar-type systems contain  gas giants \citep{Cumming2008,Mayor2011}. The formation of a gas giant requires a solid core mass that exceeds $10 \Me$ \citep{Pollack1996} before the gas disk is depleted.  Disk migration is an efficient way to transport embryos to construct such cores, but it also depends on detailed disk models \citep{Cossou2014,coleman2014,Liu2015}. On the other hand, (sub)millimeter dust observations suggest the existence of massive pebble reservoirs ($\sim 100 \Me$) in the outer region of  disks during the gas-rich phase \citep{Ricci2010,Andrews2013,Ansdell2016}. In the pebble accretion scenario the inward drift of these pebbles provide the building blocks to form massive planets \citep{Lambrechts2014,Bitsch2015b}.  Here we only focus on the pebble accretion scenario model.

Once the core mass reaches the $10 \Me$, the gas accretion enters a runaway mode \citep{Pollack1996}. As a result, the surrounding embryos and planetesimals are strongly perturbed by the sudden increase of planet's gravity \citep{Zhou2007b,Raymond2017}. These bodies could be scattered outward during close encounters. They subsequently migrate inward either due to the type I migration (embryos) or aerodynamic gas drag (planetesimal), and could be capture into the $2$:$1$ mean motion resonance with this giant planet \citep{Zhou2007b}.  Here, we only focus on the largest embryo among these bodies because it has the highest chance to grow into a secondary planet.  The embryo's  eccentricity will be excited at the resonant location.  Meanwhile, it can accrete pebbles that drift from the outer part of the disk. With the pebble accretion prescription of \se{expression}, we conduct N-body simulations to simulate the embryo's orbital and mass evolution. We explore how massive the pebble disk must be to produce a secondary planet.

We restrict our problem to a quiescent (low turbulent) disk such that pebbles are settled into the disk midplane. The accretion is thus in the $2$D regime (pebble scale height $H_{\rm P}$ is smaller than impact radius $b_{\rm set}$). The influence of disk turbulence and the prescription of $3$D pebble accretion will be presented in Paper II. 
We use the pebble generation model derived by \cite{Lambrechts2014}.
The adopted gas surface density, the disk aspect ratio are \citep{Lambrechts2014}
\begin{equation} 
    \Sigmag = 500 r_{\rm AU}^{-1} {\rm \ gcm^{-2}}, \ \hg = 0.033 r_{\rm AU}^{1/4},
\end{equation} 
where $r_{\rm AU}= r/(1 \rm \ AU)$,  the headwind prefactor $\eta =1.5 \times 10^{-3} r_{\rm AU}^{1/2}$  and the dust-to-gas ratio is $1\%$. 
 
The formation of the first gas giant may already take a few Myr, depending on the opacity in its envelope  \citep{Movshovitz2010}.  Thus,  $\dot M_{\rm disk} = 5 \times 10^{-5} \Me/yr$ is the pebble mass flux estimated from  \cite{Lambrechts2014} (assuming $t=5 \times 10^6$ yr in their Eq. (14)).   The  pebble aerodynamal size is $\taus = 0.05$ based on their Eqs. (20) and (25).  The above $\taus$ is consistent with advanced dust coagulation model \citep{Birnstiel2012} as well as disk observations \citep{Perez2015,Tazzari2016}.

A gas giant of  $1 M_{\rm J}$  is initialized at $1 \AU$.  We ignore the migration of the gas giant, but  do consider the eccentricity damping of the disk gas, which operates on a timescale of $\taue = 10^{3} \yr$ \citep{Bitsch2013}. Since  the Jupiter mass planet is much more massive than the  embryo, the eccentricity of the gas giant remains low even without any gas damping. We find that the specific choice of $\taue$ for the giant planet does not affect our simulation results. 
A $0.1 \Me$ embryo is initialized at a period ratio of $2.1$ (just  exterior to the $2$:$1$ resonance) with the inner giant planet.

We consider two scenarios: (i) a resonant case where the embryo experiences type I migration and is captured in resonance; and (ii) a non-resonant case without type I migration but including eccentricity damping for a comparison. 
For the resonant case, we implement the additional accelerations of the planet-disk interaction and eccentricity damping into the N-body code based on \citet{Papaloizou2000} and \citet{Cresswell2008}: $\bm{a}_{\rm m} = - \bm{v_{\rm p}}/ \taum$ and $\bm{a}_{\rm e} = - 2 (\bm{v_{\rm p}} \bm\cdot \bm{r_{\rm p}})  \bm{r_{\rm p}} /r_{\rm p}^2 \taue$. The embryo's migration timescale $\taum$ and the eccentricity damping timescale $\taue$ are adopted from Eqs. (11) and (13) in \cite{Cresswell2008}. These expressions  take into account the supersonic regime when $\ep \gtrsim \hg$. 
Thus, for the resonant case we expect that the embryo will migrate and get trapped into the $2$:$1$ resonance while for the comparison case the embryo will remain at the original orbit.    
It should be noted that we do not consider the inclinations of both planets. Since the inclination damping timescale is much shorter compared to the migration timescale \citep{Bitsch2011}, their inclinations would be soon damped down, restricting both planets to coplanar orbits. Thus, the above simplification is appropriate. 

The gas giant also opens a gap in the disk gas with a typical width of Hill radius. The depletion of the gas near the planet also changes the gas pressure gradient, resulting in the truncation of drifting pebbles. Therefore, a wider dust gap can be formed exterior to the gas gap. Hydrodynamic simulations show that the width of the dust gap produced by a Jupiter mass planet is  less than the distance between the giant planet and its $2$:$1$ resonance \citep{Zhu2012}. Therefore, embryos at the $2$:$1$ resonance can still accrete pebbles that drift from the outer region of the disk. \footnote{ In a very low viscous disk the the gap opened by the gas giant may be wider than its $2$:$1$ resonance. Our model discussed in this section would not apply in that case. }

\begin{figure}[t]
\includegraphics[scale=0.5, angle=0]{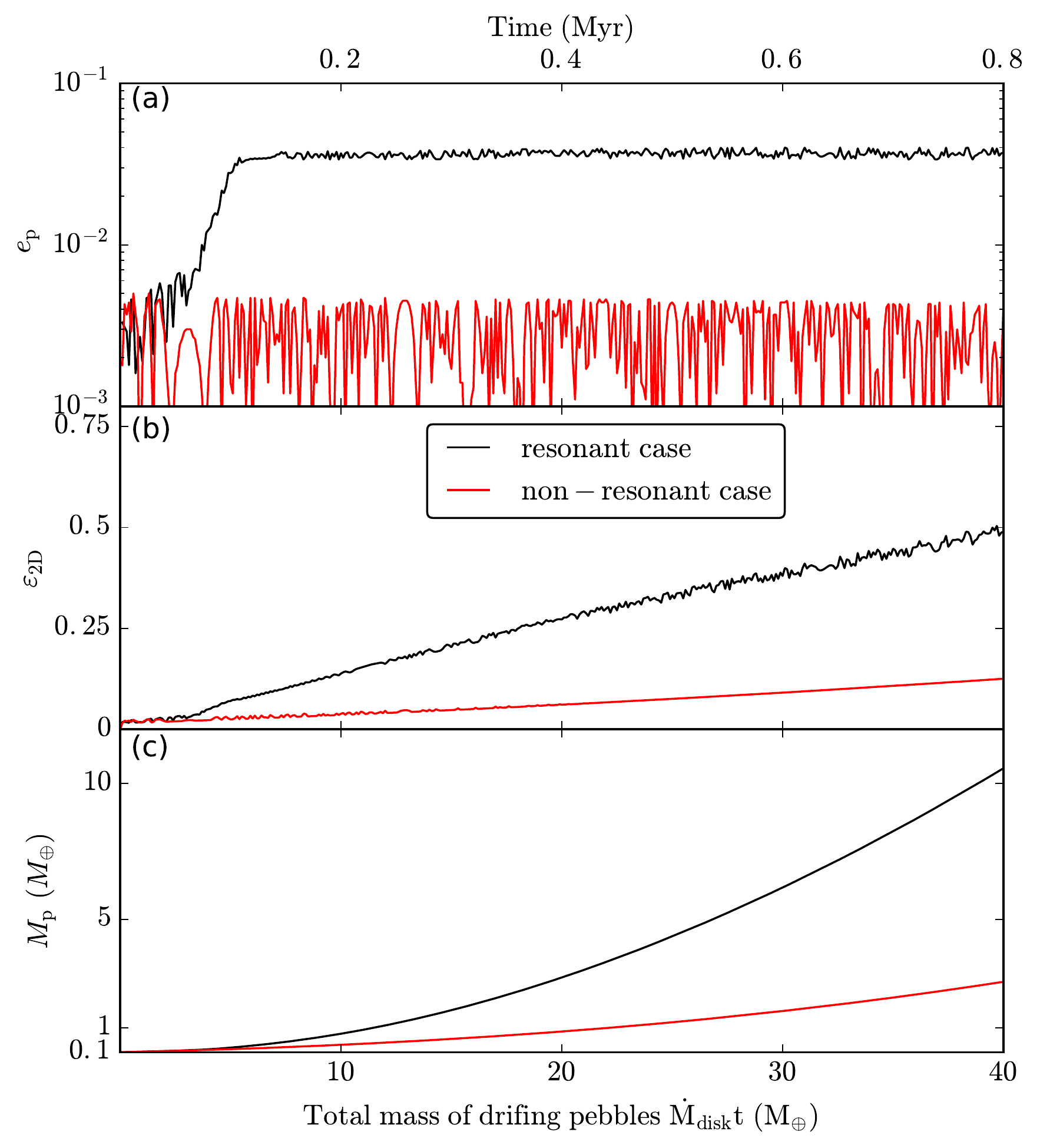}
\caption{Eccentricity (a), pebble accretion efficiency (b) and mass of the embryo (c) vs total mass of drifting pebbles (lower x axis) and time (upper x axis).
        The black line corresponds to the resonant case with type I migration and resonance trapping, while the red line represents the non-resonant, comparison case that the embryo without type I migration moves on a nearly circular orbit. 
} 
\label{fig:fsp}
\end{figure}

\fg{fsp} shows  a) the eccentricity, b) the mass growth, and c) the pebble accretion efficiency of the embryo as functions of the total mass of drifting pebbles (lower x-axis) and time (upper x-axis). 
For the resonant case, the embryo gets trapped into the $2$:$1$ resonance where  the eccentricity is gradually excited (black line in \fg{fsp}a). Meanwhile, the gas damps the eccentricity and circularizes the orbit. An equilibrium eccentricity is achieved by balancing  gas damping and resonant excitation. The equilibrium eccentricity ($e_{\rm eq}$) can be analytically derived by solving Lagrange's planetary equations with migration and eccentricity damping (see details in \cite{Teyssandier2014}). Due to the high mass ratio of the giant planet and the embryo, from their Eq. (46) $e_{\rm eq}$ can be simplified as $e_{\rm eq} \simeq \sqrt{\taue/\taum} \simeq \hg$.
In \fg{fsp}a we also find $e_{\rm eq} \simeq 0.035$, consistent with the above analysis. However, the embryo in the comparison case without migration has a nearly circular orbit  ($e_{\rm p} < 4 \times 10^{-3}$; red line in \fg{fsp}a). It is non-zero due to weak, secular perturbations of the inner gas giant. 
 

We find in \fg{fsp}b that after $3\times 10^{4}$ yr  the pebble accretion efficiency in the resonant case is higher than the non-resonant, comparison case.   As was already seen from \fg{default}, this is again because  for the resonant case the moderate eccentricity in the $2$:$1$ resonance facilitates a higher pebble accretion rate.  
After that,  a higher eccentricity in the resonant case increases the pebble accretion efficiency, leading to a more massive planet. Because of its higher mass, it  accretes pebbles more efficiently, promoting an even higher pebble accretion efficiency. 
That is why in \fg{fsp}b both efficiencies increase with time, but the efficiency in the resonant case (black) is higher than  the comparison case (red). 
As a result, in \fg{fsp}c the resonant embryo grows more rapidly than the non-resonant planet. And this mass difference increases with time.

In \fg{fsp}c we see that the final mass of the embryo depends on the total amount of pebbles available in the outer disk. Since the formation of the giant planet already consumes large amounts of pebbles, the amount of pebbles left behind that feed the secondary planet should be limited.  For instance, if the total mass of drifting pebbles beyond the planet's orbit is limited to $20 \Me$, the embryo at resonance can grow to a $3 \Me$ super-Earth, whereas the embryo in a non-resonant orbit will only reach $ 1 \Me$.
Similarly, if the total pebble mass is limited to $40 \Me$, the embryo in resonance can grow into a planet with $11 \Me$, whereas the non-resonant embryo only attains $3 \Me$. For the resonant case, then, the massive core is able to initiate the rapid gas accretion and grow into a giant planet. However, for the non-resonant case, the slow growth of the embryo results in a final super-Earth mass planet.

To conclude, an embryo at resonance accretes pebbles more efficiently and grows faster than neighboring embryos that move on nearly circular orbits. A secondary planet  therefore preferentially forms at the resonance location. Radial velocity surveys show that multiple gas giants  pile up  the $2$:$1$ resonance \citep{Wright2011}, supporting our hypothesis.

%% file: conclusion.tex
\section{Conclusions}
\label{sec:conclusion}
 
Pebble accretion is an important mechanism that drives planet growth by accreting millimeter to centimeter size particles in gas-rich disks.  Formed in the outer regions of disks, pebbles drift inward due to the aerodynamic drag from the disk gas.  
When pebbles cross the orbit of a planet, a fraction of them will be accreted by the planet. In this paper, we have calculated the efficiency of pebble accretion in the $2$D limit by conducting a series of numerical integrations of the pebble's equation of motion in both the local (co-moving) frame and the global (heliocentric) frame.

The key findings of this study are the following:
\begin{enumerate}
    \item The results of the local and global methods are generally consistent.  However, the global method more accurately simulates the pebble-planet interaction, due to the fact that  it accounts for curvature effects and models the dynamics of the pebble properly.  Since the equations of motions are linearized in the local frame,  the local method tends to overestimate $\varepsilon$ when the planet is more massive than a few Earth masses, or when the aerodynamic size of the pebble is larger than  $1$ (\se{compare}). 
    \item  We find that the 2D efficiency ($\varepsilon_\mathrm{2D}$) is a function of the planet eccentricity. A planet will accrete pebbles at a higher efficiency once the eccentricity velocity is higher than the relative velocity obtained for a circular orbit planet ($v_{\rm ecc}> v_{\rm cir}$). The pebble accretion efficiency then increases with eccentricity. However, $\varepsilon_\mathrm{2D}$ drops quickly when the eccentricity becomes too large for encounters to satisfy the settling conditions. The accretion therefore transitions to the ballistic regime. Planets with moderate eccentricities ($\ep \sim10^{-2} - 0.1$) accrete pebbles at rates a factor of $3-5$ higher than planets on circular orbits (\se{ecc}).
    \item We have obtained a recipe for the pebble accretion efficiency $\varepsilon_\mathrm{2D}$ as functions of the planet eccentricity ($\ep$), the mass ratio of the planet to the star ($\qp$), the disk headwind prefactor ($\eta$), and the aerodynamic size of the pebble ($\taus$).  Consistent with previous work,  the efficiency increases with the planet-to-star mass ratio, and decreases with both the headwind velocity and the pebble size. An analytical fit expression of $\varepsilon_\mathrm{2D} (\ep, \qp, \eta, \taus)$ is derived from our simulations (\se{expression}). Such a recipe can be readily implemented into N-body codes to study the long-term growth and evolution of planetary systems. 
    \item  In the 2D limit  embryos trapped in resonance and on eccentric orbits grow faster than those on circular orbits. Therefore, the secondary planet formation occurs preferentially at resonances with the first giant planet. 
    \end{enumerate}

In this work we have focused on the 2D limit where all pebbles are in the midplane. However,  disk turbulence may lift small  pebbles from the midplane. In addition, the inclinations of the embryos/planetesimals can be excited  by their mutual gravitational interactions. Under these circumstances, the pebble accretion efficiency is not $\varepsilon_\mathrm{2D}$, but rather involves effects related with the planet's inclination, the pebble accretion radius, and the scale height of the pebble layer \citep{Ormel2012,Morbidelli2015,Levison2015a,Levison2015b,Xu2017}. These 3D effects will be addressed in the subsequent paper (Paper II).

%% file: appendix.tex
\appendix
\section{Derivation of the accretion efficiencies}
\label{sec:appendix}

\subsection{Settling regime -- planar limit (2D)}
\label{sec:appendixset}
\subsubsection{General expression}
\label{sec:appendixge}
We use physical, order-of-magnitude arguments to derive the scaling relations of the pebble accretion efficiency (see also \cite{Ormel2010,Ormel2012,Lambrechts2012,Guillot2014,Morbidelli2015,Ida2016}). These expressions are complemented by prefactors which we obtain from the numerical simulations.  
 
There are two key requirements for pebble accretion (see \cite{Ormel2017} for a review):
\begin{enumerate}
    \item During a pebble-planet encounter, the time a pebble settles onto the planet is shorter than its encounter time, $t_{\rm set} < t_{\rm enc}$ such that the pebble is able to hit the planet. 
     \item The stopping time is shorter than the encounter time, $t_{\rm s} < t_{\rm enc}$. Otherwise, the gas drag is not important during this encounter.
\end{enumerate}

\begin{figure}[t]
      \includegraphics[width=9cm,height=7cm,keepaspectratio]{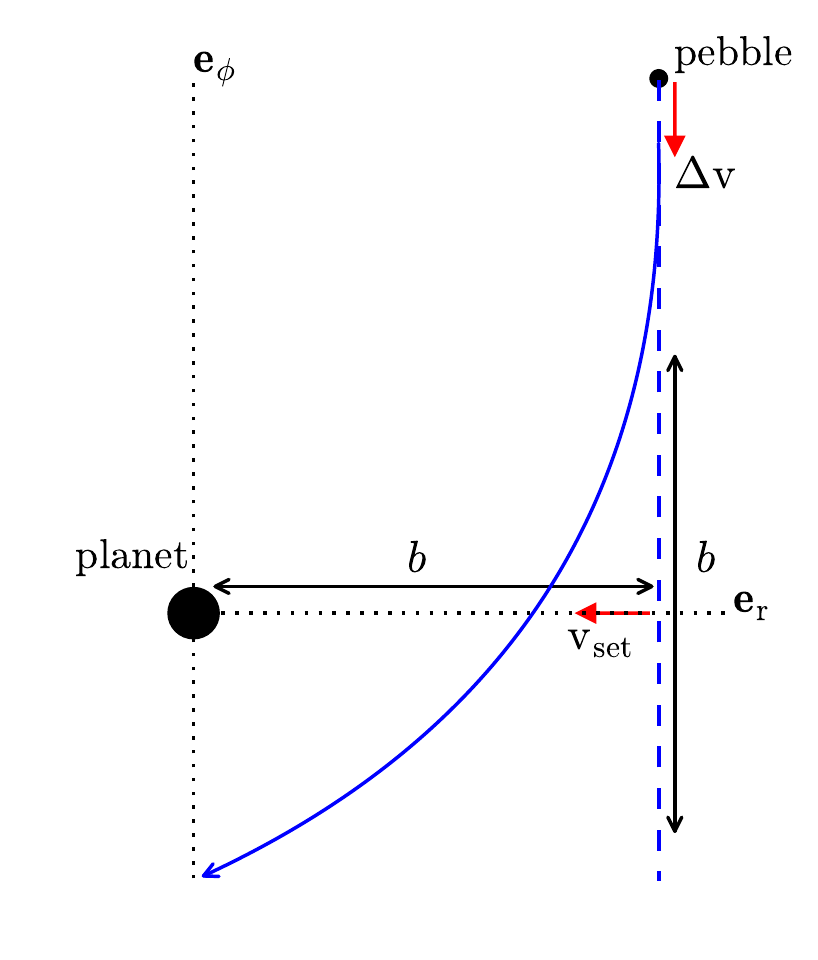}
       \caption{
           Sketch  illustrating the pebble-planet encounter in a local frame where the pebble is co-moving with the central planet. The approach velocity is given by the relative velocity between these two objects ($\Delta v$). The perturbed and unperturbed trajectories are represented in the blue solid and dashed lines.  The important timescales are: the pebble-planet encounter time $t_{\rm enc} = b/\Delta v$, and  the settling time $t_{\rm set} = b/v_{\rm set}$, where $v_{\rm set}$ is the sedimentation velocity when the planet's gravity equals to the gas drag.
     } 
\label{fig:appendix}
\end{figure} 
 
The failure of either criterion implies that accretion is not in the settling regime.  
In \fg{appendix}, the settling velocity of a pebble is obtained from balancing the planets' gravitational force with the gas drag force,  $v_{\rm set} = (GM_{\rm p}/b^2) t_{\rm s}$ where $b$ is the impact parameter.  The settling time is defined as $t_{\rm set} = b/v_{\rm set}$ while the encounter time approximates as  $t_{\rm enc} \simeq b/\Delta v$, where $\Delta v$ is the relative velocity between the planet and the pebble.   
Therefore, from criterion (1) the accretion radius of the planet in the settling regime, $b_{\rm set}$ can be written as 
\begin{equation} 
   b_{\rm set} \sim \sqrt{\frac{GM_p t_s}{\Delta v}}.
   \label{eq:reff1}
\end{equation} 
Equivalently, criterion (1) can also be obtained from the gravitational deflection time $t_g$ \citep{Perets2011,Lambrechts2012,Baruteau2016}.  For settling $t_g$ must be shorter than the stopping time, $t_{\rm g} < t_{\rm s}$. Here $t_{\rm g}= \Delta v/(G\Mp/b^2)$ is the time for deflecting the approaching velocity $\Delta v$ by the gravity of the planet.
      
In the coplanar (2D) accretion case, the mass flux of pebbles accreted by the planet is
\begin{equation} 
    \dot M_{\rm PA,2D} \sim 2 b_{\rm set} \Delta v \Sigmap \sim  2 \sqrt{G\Mp t_s \Delta v} \Sigmap.
    \label{eq:mdotpeb2d}
\end{equation} 
According to the definition, the pebble accretion efficiency for  a circular orbit planet  is given by
\begin{equation}
  \varepsilon_{\rm 0,2D} = \frac{\dot M_{\rm PA}}{\dot M_{\rm disk}} \sim \frac{   \sqrt{GM_p t_{\rm s} \Delta v} (1 + \taus^2)    }{2 \pi \taus  \eta  v_{\rm K} r }.
  \label{eq:eff0}
\end{equation}
Rewritten in terms of dimensionless quantities, this expression becomes
\begin{equation}
\varepsilon_{\rm 0,2D} =
A_{\rm 2D} \sqrt{\frac{\qp }{\taus \eta^2} \left( \frac{\Delta v}{v_{\rm K}} \right) },
\label{eq:oneefficiency0a}
\end{equation}
where $A_\mathrm{2D} =0.32$ is a fitting factor.  In the above expression, the factor $(1 + \taus^2)$ has been omitted since we focus on pebbles with $\taus <1$. In addition, dimensionless quantities are used  
\begin{equation}
    \qp \equiv \frac{M_{ \rm p}}{M_\star}= 3\times 10^{-6} \left(\frac{\Mp}{1 \Me} \right); \   
    \eta \equiv \frac{v_{\rm hw}}{v_{\rm K}} = 10^{-3} \left(\frac{\vhw}{30 \rm  \ m/s}\right)
\end{equation}

Since the pebble accretion criterion (2) breaks down when the encounter time is  shorter than the stopping time,  a transition velocity $v_\ast$  can be calculated by $t_{\rm enc} = t_{\rm s}$
\begin{equation}
v_{\ast} = (\qp /\taus)^{1/3} \vk
\label{eq:vtrans}
\end{equation}
When $\Delta v$ approaches $v_\ast$, the accretion transitions from the setting regime to the ballistic regime. We adopt an exponential decay function to fit such a transition\footnote{The form of the modulation factor is very similar to \citet{Ormel2010}, \citet{Ormel2012}, and \citet{Visser2016}, but the numerical factors are slightly different from these works.}
\begin{equation} 
 f_{\rm set} = \exp[ -0.5(\Delta v/v_{\ast})^{2} ].
\end{equation}
When $\Delta v \ll v_{\ast} $, accretion operates in the settling regime and $f_{\rm set} = 1$; on the other hand, when $\Delta v \gg v_{\ast} $, accretion is in the ballistic regime and $f_{\rm set}=0$. Therefore, the general expression of the accretion efficiency in the settling regime reads
 \begin{equation}
\varepsilon_{\rm set,2D}
  = A_{\rm 2D} \sqrt{\frac{\qp }{\taus \eta^2} \left( \frac{\Delta v}{v_{\rm K}} \right) } f_\mathrm{set},
\label{eq:efficiency1}
\end{equation}

\subsubsection{Circular \& Eccentric cases}
\label{sec:appendixecc}
For a planet on a circular orbit, its relative velocity $\Delta v$ is in principle the sum of the headwind velocity, $v_{\rm hw} \simeq \eta \vk$, and the Keplerian shear velocity between the planet and the pebble, $ v_{\rm sh} \sim \Omegak b_{\rm set}$  (\fg{frames}). In the literature, the pebble accretion in the circular {($\ep=0$) limit is classified into two regimes depending on which contribution dominates $\Delta v$: the headwind regime (Bondi regime) or the shear regime (Hill regime) \citep{Lambrechts2012,Guillot2014}. Note that  the shear velocity ($v_{\rm sh} \propto b_{\rm set}$) increases with $\taus$ and $\Mp$. Therefore, for given pebble size and disk properties (same $\taus$ and $\vhw$), planetesimals and low-mass planets tend to be in the headwind regime, whereas massive planets would be in the shear regime. Equivalently, expressed in terms of $\taus$, accretion of small $\taus$ pebbles takes place in the headwind regime whereas accretion of large $\taus$ pebbles will be in the shear regime (as seen from \fg{diff}).


When $\vhw \simeq v_{\rm sh}$ (or $\varepsilon_{\rm 0,hw} \simeq \varepsilon_{\rm 0,sh} $),  a transition mass ratio for the above two regimes is defined as $\qc = \eta^3/\taus$.  

We construct an expression for the relative velocity $\Delta v$ to be used in \eq{eff0}, which combines both the headwind and the shear regimes. For a circular orbit planet, we define  $\Delta v = v_{\rm cir}$ with
\begin{equation} 
   v_{\rm cir} =  \left[1 + a_{\rm hw/sh} \left(\frac{\qp}{\qc}\right) \right]^{-1} v_{\rm hw} + v_{\rm sh},
   \label{eq:vcir2}
\end{equation}
where $v_{\rm hw} = \eta \vk$ and $v_{\rm sh} =0.52 (\qp \taus)^{1/3} \vk$. Numerically, we fit $a_{\rm hw/sh}= 5.7$. \Eq{vcir2} provides a smooth transition at the boundary, ensuring that $ v_{\rm cir}= v_{\rm hw}$ for $\qp \ll \qc$ and $ v_{\rm cir}= v_{\rm sh}$ for $\qp \gg \qc$.  

When a planet is on a eccentric orbit, the relative velocity in addition includes an eccentricity contribution due to the elliptic orbit of the planet.  $v_{\rm ecc}$ is  the eccentric velocity of the planet relative to its circular Keplerian value. 
Therefore, in \eq{efficiency1}
 \begin{equation}
\Delta v =  \rm {max} (v_{\rm cir}, v_{\rm ecc}).
\end{equation}
We fit  $v_{\rm ecc} = 0.76 \ep \vk$ from simulations, consistent with the analysis of \cite{Guillot2014}. 
When the eccentricity is large that  $v_{\rm ecc}> v_{\rm cir}$, the accretion is in the eccentricity regime instead of the shear/headwind regime.

\subsection{Settling regime -- 3D limit}
For completeness, we also include the expression of  3D accretion efficiency here.  
The mass flux of pebbles accreted by the planet is 
\begin{equation} 
    \dot M_{\rm set,3D} =  b_{\rm set}^2 \Delta v \rho_{\rm P} \sim  G\Mp t_s  \rho_{\rm P},
    \label{eq:mdotpeb3d}
\end{equation} 
where  $\rho_{\rm P}  = \Sigma_{\rm P}/( \sqrt{2 \pi}r_{\rm p} h_{\rm P})$ is the volume density of the pebbles and $h_{\rm P}$ is the aspect ratio of the pebble layer.  The  expression of the pebble accretion efficiency in the 3D limit reads 
\begin{equation}
    \varepsilon_\mathrm{set,3D} = A_{\rm 3D} \frac{\qp}{\eta h_{\rm P}} f_\mathrm{set}^2
\end{equation}
where  $A_{\rm 3D}$ will be calculated in Paper II.
Note that in the  expression for $\varepsilon_\mathrm{set, 3D}$ $\Delta v$ only appears in $f_\mathrm{set}$.

\subsection{Ballistic regime}
  
In the ballistic regime, the accretion radius is significantly reduced due to the lack of gas drag, resulting in a drop of the accretion efficiency (see \fg{default}). 
The ballistic accretion radius now reads \citep{Safronov1972}
 \begin{equation} 
  b_{\rm bal} = R_{\rm p} \sqrt{\left( \frac{v_{\rm esc}}{\Delta v }\right)^2 +1 }.
  \label{eq:reffbl}
\end{equation}
The accretion efficiency in the $2$D ballistic regime becomes
\begin{equation} 
 \varepsilon_{\rm bal,2D} = \frac{\Delta v b_{\rm bal}  }{ 2\pi \rp  \taus \eta  v_{\rm K} }
 = \frac{ R_{\rm p} \sqrt{  {v_{\rm esc}}^2 + {\Delta v}^2 }   }{ 2\pi \rp  \taus \eta  v_{\rm K} }.
  \label{eq:effbl01}
\end{equation}
We note that when $v_{\ast} \lesssim \Delta v \lesssim v_{\rm esc}$, the accretion is in the gravitational focusing regime and  $\varepsilon_{\rm bal,2D}$ is independent of the eccentricity. On the other hand, when $\Delta v\gtrsim v_{\rm esc}$, the accretion is geometric regime. The efficiency increases with the eccentricity and $b_{\rm bal}$ is reduced to the physical radius of the planet \citep{Guillot2014}.
 
We rewrite the above formula including the ($1-f_{\rm set}$) term
\begin{equation} 
 \varepsilon_{\rm bal,2D} 
 = \frac{ R_{\rm p} } { 2\pi  \taus \eta r_{\rm p}  } \sqrt{   \frac{2 \qp r_{\rm p}}{R_{\rm p } } +\left( \frac{\Delta v}{\vk} \right)^{2}     } \left(1 - f_{\rm set}\right),
    \label{eq:effbl02}
\end{equation} 
In the 3D limit, we have

\begin{equation} 
\varepsilon_{\rm bal,3D} =
 \frac{1}  {4 \sqrt{2\pi}  \eta \taus h_{\rm P}} \left(  2\qp \frac{v_{\rm K}}{\Delta v} \frac{R_{\rm p}}{r_{\rm p}} +\frac{R_{\rm p}^2}{r_{\rm p}^2}  \frac{\Delta v}{v_{\rm K}} \right) \left( 1 -f_\mathrm{\rm set}^2 \right)
\end{equation} 

For notations of masses, velocities,  timescales, etc., see Table A.1.
\begin{table*}
    \caption{List of notations}
    \centering
    \begin{tabular}{lp{10cm}}
    \hline
    \hline
        Symbol                      & Description \\
    \hline
        $\varepsilon_{\rm 3D}$ or $\varepsilon_{\rm 2D}$                & 3D or 2D pebble accretion efficiency (probability of capture) \\
        $\varepsilon_0$               & Pebble accretion efficiency for planets on circular orbits \\
         $\varepsilon_{\rm set}$               &  Pebble accretion efficiency in the settling regime \\
         $\varepsilon_{\rm bal}$               &  Pebble accretion efficiency in the ballistic regime \\
           $A_{\rm 2D}$                    & Prefactor of pebble accretion efficiency in 2D\\
         $a_{\rm m}$               &  Acceleration for type I migration \\
         $a_{\rm e}$               &  Acceleration for the eccentricity damping \\
       $a_\mathrm{p}$              &  Planet semimajor axis  \\
          $b$               &  Impact parameter in the local frame\\
          $b_{\rm set}$               &  Accretion radius in the settling regime \\
          $b_{\rm bal}$               &  Accretion radius in the ballisitc regime \\        
          $e_\mathrm{p}$              &  Planet eccentricity  \\
           $e_\mathrm{eq}$              &  Equilibrium eccentricity at resonance  \\
           $\Omega_{\rm p}$                    & Keplerian angular velocity at planet's location\\
         $\eta$                    & Headwind prefactor   \\
         $\Sigma_{\rm gas}$              &  Gas surface density \\
         $\Sigma_{\rm P}$              &  Pebble surface density \\
         $\rho_{\rm gas}$              &  Gas volume density \\
         $\rho_{\rm P}$              &  Pebble volume density \\
              $f_{\rm set}$              & Attenuation factor    \\
         $H_\mathrm{gas}$            & Gas disk scaleheight \\
         $H_\mathrm{P}$            & Pebble disk scaleheight \\
         $h_\mathrm{gas}$            & Gas disk aspect ratio \\
         $h_\mathrm{P}$            & Pebble disk aspect ratio \\
         $M_\mathrm{p}$              &  Planet mass \\
         $M_\mathrm{p, sio}$              &  Pebble isolation mass \\
        $M_\mathrm{\star}$              &  Star mass \\       
       $N_\mathrm{hit}$              &  Number of pebbles hit onto the planet in the global simulation  \\
        $N_\mathrm{tot}$              &  Total number of pebbles initially given in the global simulation  \\
        $P$              & gas disk pressure \\
        $q_\mathrm{p}$              & Mass ratio between the planet and the star \\
        $R_\mathrm{H}$              &  Hill radius of the planet \\
        $R_\mathrm{Bondi}$              &  Bondi radius of the planet \\
        $R_\mathrm{p}$              &  Planet physical radius \\
        $r_\mathrm{p}$              &   Distance between the planet and the central star  \\
         $r$              & Distance between the pebble  and  the central pebble    \\
     $t_{\rm stop}$                    & Stopping time of the pebble\\
        $t_{\rm syn}$                    &  Synodical time of the pebble \\
        $\tau_{\rm m}$               &  Type I migration  timescale \\
        $\tau_{\rm e}$               &  Eccentricity damping  timescale \\
         $\tau_{\rm s}$                    & Dimensionless stopping time ($\tau_{\rm s}=t_\mathrm{s}\Omega_{\rm K}$) \\
        $v_{\rm esc}$               &  Planet escape velocity  \\
        $v_{\ast}$              & Transition velocity from the settling regime to the ballistic regime   \\
        $v_{\rm r}$              & Radial drift velocity of the pebble  \\
        $v_{\rm \phi}$              & Azimuthal velocity of the pebble  \\
         $v_{\rm hw}$                    & Headwind velocity of disk gas  \\
           $v_{\rm sh}$                    & Keplerian shear velocity between the planet and the pebble  \\
            $v_{\rm ecc}$                    & Relative velocity between a planet on an eccentric and a circular orbit   \\
         $v_{\rm cir}$                    & Relative velocity between the pebble and the planet on a circular orbit  \\
         $v_{\rm K}$                    & Keplerian velocity at planet's location\\
        $\dot M_{\rm PA}$              &  Pebble mass accretion rate onto the planet  \\
        $\dot M_{\rm disk}$              &  Pebble mass flux in the disk   \\
        $\Delta r_{\rm p}$              & Distance between the pebble and the planet    \\
         $\Delta v$               &  Relative velocity between the pebble and the planet \\

           \hline
    \end{tabular}
\end{table*}

%% file: main.bbl
\begin{thebibliography}{76}
\expandafter\ifx\csname natexlab\endcsname\relax\def\natexlab#1{#1}\fi

\bibitem[{{Alibert}(2017)}]{Alibert2017}
{Alibert}, Y. 2017, \aap, 606, A69

\bibitem[{{Andrews} {et~al.}(2013){Andrews}, {Rosenfeld}, {Kraus}, \&
  {Wilner}}]{Andrews2013}
{Andrews}, S.~M., {Rosenfeld}, K.~A., {Kraus}, A.~L., \& {Wilner}, D.~J. 2013,
  \apj, 771, 129

\bibitem[{{Ansdell} {et~al.}(2017){Ansdell}, {Williams}, {Manara}, {Miotello},
  {Facchini}, {van der Marel}, {Testi}, \& {van Dishoeck}}]{Ansdell2017}
{Ansdell}, M., {Williams}, J.~P., {Manara}, C.~F., {et~al.} 2017, \aj, 153, 240

\bibitem[{{Ansdell} {et~al.}(2016){Ansdell}, {Williams}, {van der Marel},
  {Carpenter}, {Guidi}, {Hogerheijde}, {Mathews}, {Manara}, {Miotello},
  {Natta}, {Oliveira}, {Tazzari}, {Testi}, {van Dishoeck}, \& {van
  Terwisga}}]{Ansdell2016}
{Ansdell}, M., {Williams}, J.~P., {van der Marel}, N., {et~al.} 2016, \apj,
  828, 46

\bibitem[{{Barenfeld} {et~al.}(2016){Barenfeld}, {Carpenter}, {Ricci}, \&
  {Isella}}]{Barenfeld2016}
{Barenfeld}, S.~A., {Carpenter}, J.~M., {Ricci}, L., \& {Isella}, A. 2016,
  \apj, 827, 142

\bibitem[{{Baruteau} {et~al.}(2016){Baruteau}, {Bai}, {Mordasini}, \&
  {Molli{\`e}re}}]{Baruteau2016}
{Baruteau}, C., {Bai}, X., {Mordasini}, C., \& {Molli{\`e}re}, P. 2016, \ssr,
  205, 77

\bibitem[{{Birnstiel} {et~al.}(2010){Birnstiel}, {Dullemond}, \&
  {Brauer}}]{Birnstiel2010}
{Birnstiel}, T., {Dullemond}, C.~P., \& {Brauer}, F. 2010, \aap, 513, A79

\bibitem[{{Birnstiel} {et~al.}(2012){Birnstiel}, {Klahr}, \&
  {Ercolano}}]{Birnstiel2012}
{Birnstiel}, T., {Klahr}, H., \& {Ercolano}, B. 2012, \aap, 539, A148

\bibitem[{{Bitsch} {et~al.}(2013){Bitsch}, {Crida}, {Morbidelli}, {Kley}, \&
  {Dobbs-Dixon}}]{Bitsch2013}
{Bitsch}, B., {Crida}, A., {Morbidelli}, A., {Kley}, W., \& {Dobbs-Dixon}, I.
  2013, \aap, 549, A124

\bibitem[{{Bitsch} \& {Kley}(2011)}]{Bitsch2011}
{Bitsch}, B. \& {Kley}, W. 2011, \aap, 530, A41

\bibitem[{{Bitsch} {et~al.}(2015){Bitsch}, {Lambrechts}, \&
  {Johansen}}]{Bitsch2015b}
{Bitsch}, B., {Lambrechts}, M., \& {Johansen}, A. 2015, \aap, 582, A112

\bibitem[{{Bitsch} {et~al.}(2018){Bitsch}, {Morbidelli}, {Johansen}, {Lega},
  {Lambrechts}, \& {Crida}}]{Bitsch2018}
{Bitsch}, B., {Morbidelli}, A., {Johansen}, A., {et~al.} 2018, ArXiv e-prints

\bibitem[{{Brauer} {et~al.}(2008){Brauer}, {Dullemond}, \&
  {Henning}}]{Brauer2008}
{Brauer}, F., {Dullemond}, C.~P., \& {Henning}, T. 2008, \aap, 480, 859

\bibitem[{{Brouwers} {et~al.}(2017){Brouwers}, {Vazan}, \&
  {Ormel}}]{Brouwers2017}
{Brouwers}, M.~G., {Vazan}, A., \& {Ormel}, C.~W. 2017, ArXiv e-prints

\bibitem[{{Coleman} \& {Nelson}(2014)}]{coleman2014}
{Coleman}, G.~A.~L. \& {Nelson}, R.~P. 2014, \mnras, 445, 479

\bibitem[{{Cossou} {et~al.}(2014){Cossou}, {Raymond}, {Hersant}, \&
  {Pierens}}]{Cossou2014}
{Cossou}, C., {Raymond}, S.~N., {Hersant}, F., \& {Pierens}, A. 2014, \aap,
  569, A56

\bibitem[{{Cresswell} \& {Nelson}(2008)}]{Cresswell2008}
{Cresswell}, P. \& {Nelson}, R.~P. 2008, \aap, 482, 677

\bibitem[{{Cumming} {et~al.}(2008){Cumming}, {Butler}, {Marcy}, {Vogt},
  {Wright}, \& {Fischer}}]{Cumming2008}
{Cumming}, A., {Butler}, R.~P., {Marcy}, G.~W., {et~al.} 2008, \pasp, 120, 531

\bibitem[{{Dominik} {et~al.}(2007){Dominik}, {Blum}, {Cuzzi}, \&
  {Wurm}}]{Dominik2007}
{Dominik}, C., {Blum}, J., {Cuzzi}, J.~N., \& {Wurm}, G. 2007, in Protostars
  and Planets V, ed. {B.~Reipurth, D.~Jewitt, \& K.~Keil} (Univ. of Arizona
  Press, Tucson), 783--800

\bibitem[{{Draine}(2006)}]{Draine2006}
{Draine}, B.~T. 2006, \apj, 636, 1114

\bibitem[{{Fehlberg}(1969)}]{Fehlberg1969}
{Fehlberg}, E. 1969, NASA Technical Report

\bibitem[{{Guillot} {et~al.}(2014){Guillot}, {Ida}, \& {Ormel}}]{Guillot2014}
{Guillot}, T., {Ida}, S., \& {Ormel}, C.~W. 2014, \aap, 572, A72

\bibitem[{{G{\"u}ttler} {et~al.}(2010){G{\"u}ttler}, {Blum}, {Zsom}, {Ormel},
  \& {Dullemond}}]{Guttler2010}
{G{\"u}ttler}, C., {Blum}, J., {Zsom}, A., {Ormel}, C.~W., \& {Dullemond},
  C.~P. 2010, \aap, 513, A56

\bibitem[{{Hood}(1998)}]{Hood1998}
{Hood}, L.~L. 1998, Meteoritics and Planetary Science, 33

\bibitem[{{Ida} {et~al.}(2016){Ida}, {Guillot}, \& {Morbidelli}}]{Ida2016}
{Ida}, S., {Guillot}, T., \& {Morbidelli}, A. 2016, \aap, 591, A72

\bibitem[{{Ida} \& {Lin}(2004)}]{Ida2004a}
{Ida}, S. \& {Lin}, D.~N.~C. 2004, \apj, 604, 388

\bibitem[{{Johansen} {et~al.}(2014){Johansen}, {Blum}, {Tanaka}, {Ormel},
  {Bizzarro}, \& {Rickman}}]{Johansen2014}
{Johansen}, A., {Blum}, J., {Tanaka}, H., {et~al.} 2014, Protostars and Planets
  VI, 547

\bibitem[{{Johansen} \& {Klahr}(2005)}]{Johansen2005}
{Johansen}, A. \& {Klahr}, H. 2005, \apj, 634, 1353

\bibitem[{{Johansen} \& {Lambrechts}(2017)}]{Johansen2017}
{Johansen}, A. \& {Lambrechts}, M. 2017, Annual Review of Earth and Planetary
  Sciences, 45, 359

\bibitem[{{Johansen} {et~al.}(2015){Johansen}, {Mac Low}, {Lacerda}, \&
  {Bizzarro}}]{Johansen2015}
{Johansen}, A., {Mac Low}, M.-M., {Lacerda}, P., \& {Bizzarro}, M. 2015,
  Science Advances, 1, 1500109

\bibitem[{{Krijt} {et~al.}(2016){Krijt}, {Ormel}, {Dominik}, \&
  {Tielens}}]{Krijt2016}
{Krijt}, S., {Ormel}, C.~W., {Dominik}, C., \& {Tielens}, A.~G.~G.~M. 2016,
  \aap, 586, A20

\bibitem[{{Lambrechts} \& {Johansen}(2012)}]{Lambrechts2012}
{Lambrechts}, M. \& {Johansen}, A. 2012, \aap, 544, A32

\bibitem[{{Lambrechts} \& {Johansen}(2014)}]{Lambrechts2014}
{Lambrechts}, M. \& {Johansen}, A. 2014, \aap, 572, A107

\bibitem[{{Lambrechts} {et~al.}(2014){Lambrechts}, {Johansen}, \&
  {Morbidelli}}]{Lambrechts2014b}
{Lambrechts}, M., {Johansen}, A., \& {Morbidelli}, A. 2014, \aap, 572, A35

\bibitem[{{Lee} \& {Peale}(2002)}]{Lee2002}
{Lee}, M.~H. \& {Peale}, S.~J. 2002, \apj, 567, 596

\bibitem[{{Levison} {et~al.}(2015{\natexlab{a}}){Levison}, {Kretke}, \&
  {Duncan}}]{Levison2015a}
{Levison}, H.~F., {Kretke}, K.~A., \& {Duncan}, M.~J. 2015{\natexlab{a}}, \nat,
  524, 322

\bibitem[{{Levison} {et~al.}(2015{\natexlab{b}}){Levison}, {Kretke}, {Walsh},
  \& {Bottke}}]{Levison2015b}
{Levison}, H.~F., {Kretke}, K.~A., {Walsh}, K.~J., \& {Bottke}, W.~F.
  2015{\natexlab{b}}, Proceedings of the National Academy of Science, 112,
  14180

\bibitem[{{Lin} \& {Papaloizou}(1993)}]{Lin1993}
{Lin}, D.~N.~C. \& {Papaloizou}, J.~C.~B. 1993, in Protostars and Planets III,
  ed. E.~H. {Levy} \& J.~I. {Lunine}, 749--835

\bibitem[{{Lissauer}(1987)}]{Lissauer1987}
{Lissauer}, J.~J. 1987, \icarus, 69, 249

\bibitem[{{Liu} {et~al.}(2015){Liu}, {Zhang}, {Lin}, \& {Aarseth}}]{Liu2015}
{Liu}, B., {Zhang}, X., {Lin}, D.~N.~C., \& {Aarseth}, S.~J. 2015, \apj, 798,
  62

\bibitem[{{Matsumura} {et~al.}(2017){Matsumura}, {Brasser}, \&
  {Ida}}]{Matsumura2017}
{Matsumura}, S., {Brasser}, R., \& {Ida}, S. 2017, \aap, 607, A67

\bibitem[{{Mayor} {et~al.}(2011){Mayor}, {Marmier}, {Lovis}, {Udry},
  {S{\'e}gransan}, {Pepe}, {Benz}, {Bertaux}, {Bouchy}, {Dumusque}, {Lo Curto},
  {Mordasini}, {Queloz}, \& {Santos}}]{Mayor2011}
{Mayor}, M., {Marmier}, M., {Lovis}, C., {et~al.} 2011, ArXiv e-prints:
  1109.2497

\bibitem[{{Miura} \& {Nakamoto}(2005)}]{Miura2005}
{Miura}, H. \& {Nakamoto}, T. 2005, \icarus, 175, 289

\bibitem[{{Morbidelli} {et~al.}(2015){Morbidelli}, {Lambrechts}, {Jacobson}, \&
  {Bitsch}}]{Morbidelli2015}
{Morbidelli}, A., {Lambrechts}, M., {Jacobson}, S., \& {Bitsch}, B. 2015,
  \icarus, 258, 418

\bibitem[{{Morris} {et~al.}(2012){Morris}, {Boley}, {Desch}, \&
  {Athanassiadou}}]{Morris2012}
{Morris}, M.~A., {Boley}, A.~C., {Desch}, S.~J., \& {Athanassiadou}, T. 2012,
  \apj, 752, 27

\bibitem[{{Movshovitz} {et~al.}(2010){Movshovitz}, {Bodenheimer}, {Podolak}, \&
  {Lissauer}}]{Movshovitz2010}
{Movshovitz}, N., {Bodenheimer}, P., {Podolak}, M., \& {Lissauer}, J.~J. 2010,
  \icarus, 209, 616

\bibitem[{{Murray} \& {Dermott}(1999)}]{Murray1999}
{Murray}, C.~D. \& {Dermott}, S.~F. 1999, {Solar system dynamics}

\bibitem[{{Natta} \& {Testi}(2004)}]{Natta2004}
{Natta}, A. \& {Testi}, L. 2004, in Astronomical Society of the Pacific
  Conference Series, Vol. 323, Star Formation in the Interstellar Medium: In
  Honor of David Hollenbach, ed. D.~{Johnstone}, F.~C. {Adams}, D.~N.~C. {Lin},
  D.~A. {Neufeeld}, \& E.~C. {Ostriker}, 279

\bibitem[{{Ormel}(2017)}]{Ormel2017}
{Ormel}, C.~W. 2017, in Astrophysics and Space Science Library, Vol. 445,
  Astrophysics and Space Science Library, ed. M.~{Pessah} \& O.~{Gressel}, 197

\bibitem[{{Ormel} \& {Klahr}(2010)}]{Ormel2010}
{Ormel}, C.~W. \& {Klahr}, H.~H. 2010, \aap, 520, A43

\bibitem[{{Ormel} \& {Kobayashi}(2012)}]{Ormel2012}
{Ormel}, C.~W. \& {Kobayashi}, H. 2012, \apj, 747, 115

\bibitem[{{Ormel} \& {Liu}(2018)}]{Ormel2018}
{Ormel}, C.~W. \& {Liu}, B. 2018, ArXiv e-prints

\bibitem[{{Ormel} {et~al.}(2015){Ormel}, {Shi}, \& {Kuiper}}]{Ormel2015}
{Ormel}, C.~W., {Shi}, J.-M., \& {Kuiper}, R. 2015, \mnras, 447, 3512

\bibitem[{{Papaloizou} \& {Larwood}(2000)}]{Papaloizou2000}
{Papaloizou}, J.~C.~B. \& {Larwood}, J.~D. 2000, \mnras, 315, 823

\bibitem[{{Pascucci} {et~al.}(2016){Pascucci}, {Testi}, {Herczeg}, {Long},
  {Manara}, {Hendler}, {Mulders}, {Krijt}, {Ciesla}, {Henning}, {Mohanty},
  {Drabek-Maunder}, {Apai}, {Sz{\H u}cs}, {Sacco}, \&
  {Olofsson}}]{Pascucci2016}
{Pascucci}, I., {Testi}, L., {Herczeg}, G.~J., {et~al.} 2016, \apj, 831, 125

\bibitem[{{Perets} \& {Murray-Clay}(2011)}]{Perets2011}
{Perets}, H.~B. \& {Murray-Clay}, R.~A. 2011, \apj, 733, 56

\bibitem[{{P{\'e}rez} {et~al.}(2015){P{\'e}rez}, {Chandler}, {Isella},
  {Carpenter}, {Andrews}, {Calvet}, {Corder}, {Deller}, {Dullemond}, {Greaves},
  {Harris}, {Henning}, {Kwon}, {Lazio}, {Linz}, {Mundy}, {Ricci}, {Sargent},
  {Storm}, {Tazzari}, {Testi}, \& {Wilner}}]{Perez2015}
{P{\'e}rez}, L.~M., {Chandler}, C.~J., {Isella}, A., {et~al.} 2015, \apj, 813,
  41

\bibitem[{{Picogna} {et~al.}(2018){Picogna}, {Stoll}, \& {Kley}}]{Picogna2018}
{Picogna}, G., {Stoll}, M.~H.~R., \& {Kley}, W. 2018, ArXiv e-prints

\bibitem[{{Pollack} {et~al.}(1996){Pollack}, {Hubickyj}, {Bodenheimer},
  {Lissauer}, {Podolak}, \& {Greenzweig}}]{Pollack1996}
{Pollack}, J.~B., {Hubickyj}, O., {Bodenheimer}, P., {et~al.} 1996, \icarus,
  124, 62

\bibitem[{{Raymond} \& {Izidoro}(2017)}]{Raymond2017}
{Raymond}, S.~N. \& {Izidoro}, A. 2017, \icarus, 297, 134

\bibitem[{{Raymond} {et~al.}(2006){Raymond}, {Quinn}, \&
  {Lunine}}]{Raymond2006}
{Raymond}, S.~N., {Quinn}, T., \& {Lunine}, J.~I. 2006, \icarus, 183, 265

\bibitem[{{Ricci} {et~al.}(2010){Ricci}, {Testi}, {Natta}, {Neri}, {Cabrit}, \&
  {Herczeg}}]{Ricci2010}
{Ricci}, L., {Testi}, L., {Natta}, A., {et~al.} 2010, \aap, 512, A15

\bibitem[{{Safronov}(1972)}]{Safronov1972}
{Safronov}, V.~S. 1972, {Evolution of the protoplanetary cloud and formation of
  the earth and planets.}

\bibitem[{{Sato} {et~al.}(2016){Sato}, {Okuzumi}, \& {Ida}}]{Sato2016}
{Sato}, T., {Okuzumi}, S., \& {Ida}, S. 2016, \aap, 589, A15

\bibitem[{{Tazzari} {et~al.}(2016){Tazzari}, {Testi}, {Ercolano}, {Natta},
  {Isella}, {Chandler}, {P{\'e}rez}, {Andrews}, {Wilner}, {Ricci}, {Henning},
  {Linz}, {Kwon}, {Corder}, {Dullemond}, {Carpenter}, {Sargent}, {Mundy},
  {Storm}, {Calvet}, {Greaves}, {Lazio}, \& {Deller}}]{Tazzari2016}
{Tazzari}, M., {Testi}, L., {Ercolano}, B., {et~al.} 2016, \aap, 588, A53

\bibitem[{{Teyssandier} \& {Terquem}(2014)}]{Teyssandier2014}
{Teyssandier}, J. \& {Terquem}, C. 2014, \mnras, 443, 568

\bibitem[{{Venturini} {et~al.}(2016){Venturini}, {Alibert}, \&
  {Benz}}]{Venturini2016}
{Venturini}, J., {Alibert}, Y., \& {Benz}, W. 2016, \aap, 596, A90

\bibitem[{{Visser} \& {Ormel}(2016)}]{Visser2016}
{Visser}, R.~G. \& {Ormel}, C.~W. 2016, \aap, 586, A66

\bibitem[{{Weidenschilling}(1977)}]{Weidenschilling1977a}
{Weidenschilling}, S.~J. 1977, \mnras, 180, 57

\bibitem[{{Weidenschilling} \& {Cuzzi}(1993)}]{Weidenschilling1993}
{Weidenschilling}, S.~J. \& {Cuzzi}, J.~N. 1993, in Protostars and Planets III,
  ed. E.~H. {Levy} \& J.~I. {Lunine}, 1031--1060

\bibitem[{{Williams} \& {Cieza}(2011)}]{Williams2011}
{Williams}, J.~P. \& {Cieza}, L.~A. 2011, \araa, 49, 67

\bibitem[{{Wright} {et~al.}(2011){Wright}, {Veras}, {Ford}, {Johnson}, {Marcy},
  {Howard}, {Isaacson}, {Fischer}, {Spronck}, {Anderson}, \&
  {Valenti}}]{Wright2011}
{Wright}, J.~T., {Veras}, D., {Ford}, E.~B., {et~al.} 2011, \apj, 730, 93

\bibitem[{{Xu} {et~al.}(2017){Xu}, {Bai}, \& {Murray-Clay}}]{Xu2017}
{Xu}, Z., {Bai}, X.-N., \& {Murray-Clay}, R.~A. 2017, \apj, 847, 52

\bibitem[{{Zheng} {et~al.}(2017){Zheng}, {Lin}, \& {Kouwenhoven}}]{Zheng2017}
{Zheng}, X., {Lin}, D.~N.~C., \& {Kouwenhoven}, M.~B.~N. 2017, \apj, 836, 207

\bibitem[{{Zhou} \& {Lin}(2007)}]{Zhou2007b}
{Zhou}, J.-L. \& {Lin}, D.~N.~C. 2007, \apj, 666, 447

\bibitem[{{Zhu} {et~al.}(2012){Zhu}, {Nelson}, {Dong}, {Espaillat}, \&
  {Hartmann}}]{Zhu2012}
{Zhu}, Z., {Nelson}, R.~P., {Dong}, R., {Espaillat}, C., \& {Hartmann}, L.
  2012, \apj, 755, 6

\end{thebibliography}
